\documentclass[aps,prb,twocolumn,groupedaddress,showpacs,superscriptaddress]{revtex4-1}

\usepackage{graphicx}
\usepackage{epstopdf}
\usepackage{array}
\usepackage{amsmath}
\usepackage{amssymb}
\usepackage{graphics}
\usepackage{xcolor}

\begin{document}

\title{Small influence of magnetic ordering on lattice dynamics in TaFe$_{1.25}$Te$_3$}

\author{M. Opa\v{c}i\'c}
 \affiliation{Center for Solid State Physics and New Materials, Institute of Physics Belgrade, University of Belgrade, Pregrevica 118, 11080 Belgrade, Serbia}

\author{N. Lazarevi\'c}
 \affiliation{Center for Solid State Physics and New Materials, Institute of Physics Belgrade, University of Belgrade, Pregrevica 118, 11080 Belgrade, Serbia}

\author{D. Tanaskovi\'c}
\affiliation{Scientific Computing Laboratory, Center for the Study of Complex Systems, Institute of Physics Belgrade, University of Belgrade, Pregrevica 118, 11080 Belgrade, Serbia}

 \author{M. M. Radonji\'c}
 \affiliation{Scientific Computing Laboratory, Center for the Study of Complex Systems, Institute of Physics Belgrade, University of Belgrade, Pregrevica 118, 11080 Belgrade, Serbia}

\author{A. Milosavljevi\'c}
 \affiliation{Center for Solid State Physics and New Materials, Institute of Physics Belgrade, University of Belgrade, Pregrevica 118, 11080 Belgrade, Serbia}

 \author{Yongchang Ma}
 \affiliation{Condensed Matter Physics and Materials Science Department, Brookhaven
National Laboratory, Upton, New York 11973-5000, USA}
 \affiliation{School of Materials Science and Engineering, Tianjin University of Technology, Tianjin 300384, People's Republic of China}

\author{C. Petrovic}
\affiliation{Condensed Matter Physics and Materials Science Department, Brookhaven
National Laboratory, Upton, New York 11973-5000, USA}

\author{Z. V. Popovi\'c}
\affiliation{Center for Solid State Physics and New Materials, Institute of Physics Belgrade, University of Belgrade,
Pregrevica 118, 11080 Belgrade, Serbia}
\affiliation{Serbian Academy of Sciences and Arts, Knez Mihailova 35, 11000 Belgrade, Serbia}

\date{\today}

\begin{abstract}
Raman scattering spectra of zigzag spin chain TaFe$_{1.25}$Te$_3$ single crystal are presented in a temperature range from 80 to 300 K. Nine Raman active modes of $A_g$ and $B_g$ symmetry are clearly observed and assigned by probing different scattering channels, which is confirmed by lattice dynamics calculations. Temperature dependence of the Raman modes linewidth is mainly governed by the lattice anharmonicity. The only deviation from the conventional behavior is observed for $A_g$ symmetry modes in a vicinity of the magnetic phase transition at $T_N \approx 200$ K. This implies that the electron-phonon interaction weakly changes with temperature and magnetic ordering, whereas small changes in the spectra near the critical temperature can be ascribed to spin fluctuations.
\end{abstract}

\maketitle

\section{Introduction}

The discovery of superconductivity in La(O$_{1-x}$F$_x$)FeAs in 2008 \cite{Kamihara_JACS2008} initiated an intensive search for new iron-based superconducting materials, in order to obtain better understanding of their physical properties and the mechanism of high-$T_c$ superconductivity.\cite{Ma_PRL2012, Bao_ChinPhysLett2011, Stewart_RevModPhys2011}  Novel iron-based materials, however, are not only superconducting, but can also exhibit various types of magnetic ordering. In some cases the magnetic phase transition is continuous,\cite{Opacic_JPCM2016, Gonen_CHM2000, Lei_PRB2011, Han_PRB2012} whereas in other it is accompanied by structural changes, \cite{Fang_PRB2008,Li_PRB2009,Gnezdilov_PRB2011, Um_PRB2012, Popovic_SSC2014, Choi_PRB2008, Rahlenbeck_PRB2009} or even by a nanoscale coexistence of antiferromagnetic (AFM) and superconducting domains.\cite{Chen_PRX2011, Li_NatPhys2012, Lazarevic_PRB2012}

\begin{figure}
\centering
\includegraphics[width = 85mm]{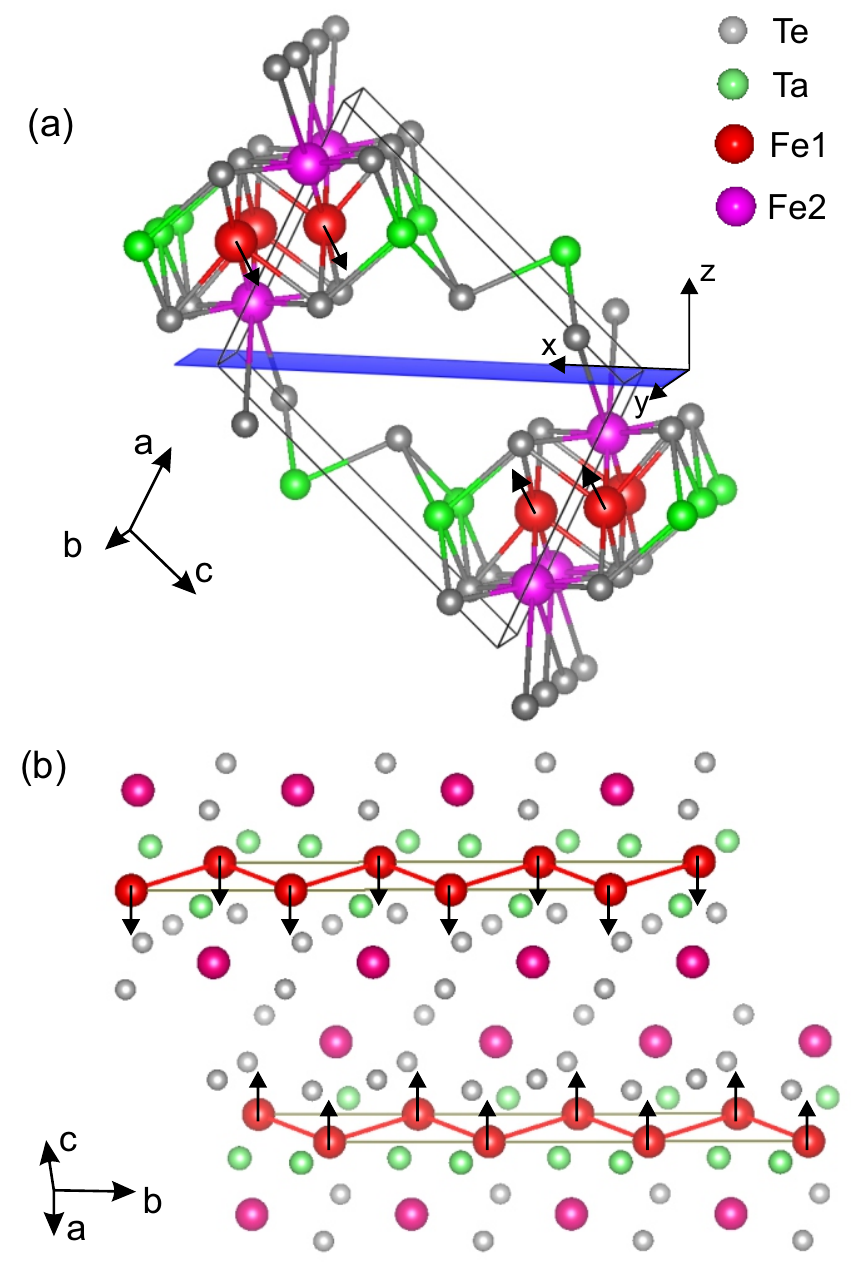}
\caption{(Color online) (a) The structure of the TaFe$_{1.25}$Te$_3$ single crystal together with the natural cleavage plane $[\bar{1}01]$. $\mathbf{x}=1/\sqrt{2}(\bar{1}0\bar{1})$ and $\mathbf{y}=(010)$ correspond to our laboratory system. (b) A view of the TaFe$_{1.25}$Te$_3$ structure along the $b$-axis. Two neighboring chains of Fe1 spins point in a parallel direction, forming a ferromagnetic zigzag chain, whereas spins of neighboring zigzag chains order antiferromagnetically. One should note that Fe2 is present with occupancy of 0.25.}
\label{fig1b}
\end{figure}

TaFe$_{1+y}$Te$_3$ was synthesized and characterized about 25 years ago.\cite{Badding_JSCS1992, Neuhausen_ZNB1993} It is a layered system consisting of FeTe chains, along the $b$-axis, separated by a Ta/Te network in-between, see Figure~\ref{fig1b}. These layers are parallel to the natural cleavage plane. There are also additional Fe ions, Fe2, randomly occupying interstitial sites.\cite{Perez_JPCB1998, Liu_PRB2011, Ke_PRB2012} TaFe$_{1+y}$Te$_3$ features anisotropic charge transport with metallic resistivity within the plane and insulating in the direction normal to the FeTe layers.\cite{Ke_PRB2012} The first study of magnetic structure implies that TaFe$_{1+y}$Te$_3$ is composed of double zigzag spin chains with antiferromagnetic ordering of Fe1 spins.\cite{Liu_PRB2011} The newest neutron diffraction measurements suggest that spin ordering within zigzag chains is ferromagnetic, whereas these zigzag chains couple antiferromagnetically,\cite{Ke_PRB2012} as shown in Fig.~\ref{fig1b}(b). However, the exact interaction mechanism is not clearly resolved.

There is a similarity between TaFe$_{1+y}$Te$_3$ and extensively studied  Fe$_{1+y}$Te compound since they are correlated bad metals which order antiferromagnetically below $T_N \approx 200$ K and 70 K, respectively,\cite{Ke_PRB2012,Li_PRB2009} both having rather large magnetic moments on Fe ions, $\approx 2$ $\mu_\mathrm{B}/\mathrm{Fe}$. TaFe$_{1+y}$Te$_3$, however, forms ferromagnetic (FM) zigzag spin chains which couple antiferromagnetically between the layers, whereas the Fe spins in Fe$_{1+y}$Te form a bicollinear AFM structure. The magnetic phase transition in Fe$_{1+y}$Te is accompanied by the structural change from a tetragonal to a monoclinic, as opposed to TaFe$_{1+y}$Te$_3$ where a continuous transition to the AFM phase is observed in thermodynamic and transport measurements. \cite{Liu_PRB2011} Just like in Fe$_{1+y}$Te, interest in spin chain and ladder materials \cite{Dagotto_RMP2013} stems not only from their block-AFM states similar to parent compounds of iron-based superconductors,\cite{Li_FP2014} but also from superconductivity. It is worth noting that spin 1/2 copper oxide ladder structures host a spin gap and superconductivity upon doping.\cite{Dagotto_PRB1992, Dagotto_SCI1996, Dagotto_RPP1999} In contrast to superconductivity in copper oxide ladder materials that was rather rare and with critical temperatures rather small when compared to highest achieved in copper square lattices,\cite{Uehara_JPSJ1996, Hisada_JPSJ2014} iron-ladder materials feature $T_c$'s similar to the highest found in Fe-based superconductors.\cite{Takahashi_NMat2014}

Raman spectra provide additional information on magnetic ordering and electron-phonon coupling. There exist several  Raman studies of the phonon spectra of iron based materials near the superconducting or magnetic phase transition.\cite{Zhang_CPB2013,opavcic:2017lattice} While no anomalies were observed in 1111 compounds,\cite{Zhang_PRB2009, Gallais_PRB2008}  the Raman spectra show anomalous behavior near the spin density wave (SDW) transition in some of the 122  and 11  compounds,\cite{Rahlenbeck_PRB2009, Gnezdilov_PRB2013, Litvinchuk_PRB2011,Choi_JPCM2010} which was ascribed to the phonon renormalization due to the opening of the SDW or superconducting gap, or to the structural transition. Large anomalies were observed also in ferromagnetic K$_x$Co$_{2-y}$Se$_2$,\cite{Opacic_JPCM2016} which was ascribed to the effect of electron-phonon coupling and spin fluctuations. Fe$_{1+y}$Te phonon spectra feature unusually large anomalies near the magnetic phase transition, as seen in sudden changes in the phonon frequencies and linewidths, due to the phonon modulation of magnetic interactions and structural phase transition.\cite{Gnezdilov_PRB2011, Um_PRB2012, Popovic_SSC2014} Therefore, it is of interest to examine lattice dynamics in the normal state of iron-spin chain and ladder materials and compare it to materials like Fe$_{1+y}$Te.  To the best of our knowledge, there are no published data on lattice dynamics of TaFe$_{1+y}$Te$_3$.

In this paper we present polarized Raman scattering spectra of TaFe$_{1.25}$Te$_3$ single crystal measured in a temperature range from 80 to 300 K. Nine out of 15 Raman active modes are observed and assigned using the selection rules for different polarization configurations and lattice dynamics calculations. In a sharp contrast to the related FeTe compound, TaFe$_{1.25}$Te$_3$ Raman spectra do not show significant changes near $T_N \approx 200$ K, which clearly indicates that the phase transition is continuous. Temperature dependence of the frequency and linewidth is conventional, driven by the anharmonicity effects, except very near $T_N$ where some of phonon lines slightly broaden which should be the consequence of spin fluctuations near the critical temperature. These results indicate very small changes in the electron-phonon coupling and in the Fermi surface in the measured temperature range.

\section{Experiment and numerical method}

Single crystals were grown using the self-flux method, as described elsewhere.\cite{Badding_JSCS1992} Raman scattering measurements were performed on a freshly cleaved $(\bar{1}01)$-oriented samples, using Jobin Yvon T64000 Raman system, equipped with a nitrogen-cooled CCD detector, in the backscattering micro-Raman configuration. The 532 nm line of a solid state laser was used as an excitation source. A microscope objective with 50$\times$ magnification was used for focusing the laser beam. All measurements were performed at low laser power, to reduce local heating of the sample. For low temperature measurements KONTI CryoVac continuous flow cryostat with 0.5 mm thick window was used. All spectra were corrected for the Bose factor. For extracting the data from the Raman spectra, phonon modes were fitted with a Lorentzian profile.

The electronic structure is calculated for stoichiometric TaFeTe$_3$ in the paramagnetic phase within the density functional theory (DFT), and  the phonon frequencies at the $\Gamma$-point are obtained within the density functional perturbation theory (DFPT),\cite{Baroni_RevModPhys2001} using the {\small QUANTUM ESPRESSO} package.\cite{Gianozzi_JPCM2009} We have used Projector Augmented Wave (PAW) pseudopotentials with Perdew-Burke-Ernzerhof (PBE) exchange-correlation functional with nonlinear core correction and Gaussian smearing of 0.01 Ry. The electron wave-function and the density energy cutoffs were 64 Ry and 782 Ry, respectively. The Brillouin zone is sampled with 8$\times$8$\times$8 Monkhorst-Pack k-space mesh. The phonon frequencies were calculated with the unit cell size taken from the experiments and the relaxed positions of atoms within the unit cell. The forces acting on individual atoms in the relaxed configuration were smaller than $10^{-4}$ Ry/a.u.

\section{Results and discussion}

\begin{figure}
\centering
\includegraphics[width = 85mm]{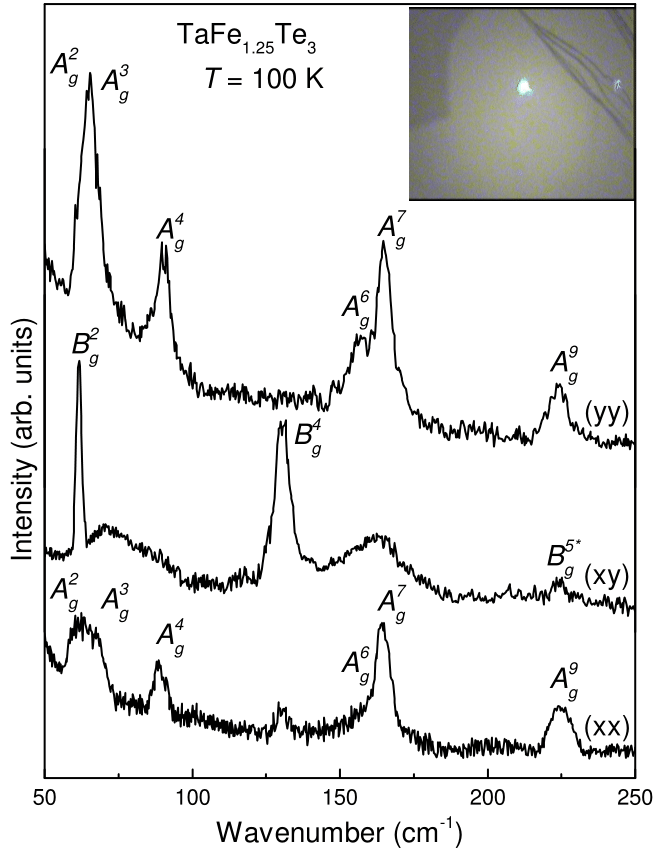}
\caption{(Color online) Polarized Raman scattering spectra of TaFe$_{1.25}$Te$_3$ single crystal measured at 100 K in various polarizations. The notation in parentheses indicates the polarization directions of the incident and scattered light according to Fig.~\ref{fig1b}(a). Inset: surface of the probed TaFe$_{1.25}$Te$_3$ single crystal.}
\label{fig2}
\end{figure}

\begin{table}[b]
\caption{Experimental fractional coordinates of TaFe$_{1.25}$Te$_3$ taken from Ref.~\onlinecite{Badding_JSCS1992}. In the parentheses are the calculated values for TaFeTe$_3$.}
\label{tab1}
\begin{ruledtabular}
\centering
\begin{tabular}{l c c c }


Atom type & $x$ & $y$ & $z$ \\ [1mm] \hline
Ta & 0.8340 (0.8331) & 0.25 & 0.3007 (0.2987) \\ [1mm]
Fe1 & 0.6147 (0.6223) & -0.25 & 0.0890 (0.0988) \\ [1mm]
Fe2 & 0.7686 & 0.25 & -0.0047 \\ [1mm]
Te1 & 0.4392 (0.4326) & 0.25 & 0.1860 (0.1637) \\ [1mm]
Te2 & 0.9835 (0.9842) & -0.25 & 0.1589 (0.1584) \\ [1mm]
Te3 & 0.2179 (0.2192) & 0.25 & 0.4970 (0.5028) \\ [1mm]

\end{tabular}
\end{ruledtabular}
\end{table}

\begin{table*}[t]
\caption{Upper panel: atomic species (all of them are at $2e$ Wyckoff positions) and the contribution of the each atom to the $\Gamma$-point phonons, the corresponding Raman tensors for the TaFeTe$_3$ single crystal ($P2_1/m$ space group).\cite{Porto:1981} Lower panel: the calculated (for the stoichiometric TaFeTe$_3$) and experimental phonon energies at 100 K (for the TaFe$_{1.25}$Te$_3$ single crystal).}
\label{tab2}
\begin{ruledtabular}
\centering
\resizebox{\linewidth}{!}{%
\begin{tabular}{c c c c}
\multicolumn{2}{c} {Atoms} & \multicolumn{2}{c} {Irreducible representations} \\

\cline{1-4} \\[-0.5em]

\multicolumn{2}{c} {Ta, Fe1, Te1, Te2, Te3} & \multicolumn{2}{c} {2$A_g$+$A_u$+$B_g$+2$B_u$} \\ [1mm] \hline \\[-1.0em]

\multicolumn{4}{c}{Raman tensors} \\

& $\hat{R}_{A_{g}} = \left(\begin{array}{ccc} a & 0 & d \\ 0 & b & 0 \\ d & 0 & c \end{array}\right)$ & $\hat{R}_{B_{g}} = \left(\begin{array}{ccc} 0 & e & 0 \\ e & 0 & f \\ 0 & f & 0 \end{array}\right)$ & \\[1mm] \hline \\[-1.0em]


\multicolumn{2}{c} {Raman active} & \multicolumn{2}{c} {Infrared active} \\ [+1.0em]

Symmetry & \begin{tabular}{>{\centering\arraybackslash}m{2.0cm} >{\centering\arraybackslash}m{2.0cm}} Calc.(cm$^{-1}$) & Exp.(cm$^{-1}$) \end{tabular} & Symmetry & \begin{tabular}{>{\centering\arraybackslash}m{2.0cm} >{\centering\arraybackslash}m{2.0cm}} Calc.(cm$^{-1}$) & Exp.(cm$^{-1}$) \end{tabular} \\ \hline \\[-1.0em]

$A_g^1$ & \begin{tabular}{>{\centering\arraybackslash}m{2.0cm} >{\centering\arraybackslash}m{2.0cm}} 36.2 & - \end{tabular} & $A_u^1$ & \begin{tabular}{>{\centering\arraybackslash}m{2.0cm} >{\centering\arraybackslash}m{2.0cm}} 42.8 & - \end{tabular} \\
$B_g^1$ & \begin{tabular}{>{\centering\arraybackslash}m{2.0cm} >{\centering\arraybackslash}m{2.0cm}} 43.8 & - \end{tabular} & $B_u^1$ & \begin{tabular}{>{\centering\arraybackslash}m{2.0cm} >{\centering\arraybackslash}m{2.0cm}} 54.9 & - \end{tabular} \\
$B_g^2$ & \begin{tabular}{>{\centering\arraybackslash}m{2.0cm} >{\centering\arraybackslash}m{2.0cm}} 57.9 & 61.6 \end{tabular} & $B_u^2$ & \begin{tabular}{>{\centering\arraybackslash}m{2.0cm} >{\centering\arraybackslash}m{2.0cm}} 94.4 & - \end{tabular} \\
$A_g^2$ & \begin{tabular}{>{\centering\arraybackslash}m{2.0cm} >{\centering\arraybackslash}m{2.0cm}} 63.8 & 62.3 \end{tabular} & $A_u^2$ & \begin{tabular}{>{\centering\arraybackslash}m{2.0cm} >{\centering\arraybackslash}m{2.0cm}} 101.4 & - \end{tabular} \\
$A_g^3$ & \begin{tabular}{>{\centering\arraybackslash}m{2.0cm} >{\centering\arraybackslash}m{2.0cm}} 75.3 & 68.5 \end{tabular} & $B_u^3$ & \begin{tabular}{>{\centering\arraybackslash}m{2.0cm} >{\centering\arraybackslash}m{2.0cm}} 111.3 & - \end{tabular} \\
$A_g^4$ & \begin{tabular}{>{\centering\arraybackslash}m{2.0cm} >{\centering\arraybackslash}m{2.0cm}} 104.4 & 90 \end{tabular} & $A_u^3$ & \begin{tabular}{>{\centering\arraybackslash}m{2.0cm} >{\centering\arraybackslash}m{2.0cm}} 131.1 & - \end{tabular} \\
$B_g^3$ & \begin{tabular}{>{\centering\arraybackslash}m{2.0cm} >{\centering\arraybackslash}m{2.0cm}} 105.1 & - \end{tabular} & $B_u^4$ & \begin{tabular}{>{\centering\arraybackslash}m{2.0cm} >{\centering\arraybackslash}m{2.0cm}} 143.2 & - \end{tabular} \\
$A_g^5$ & \begin{tabular}{>{\centering\arraybackslash}m{2.0cm} >{\centering\arraybackslash}m{2.0cm}} 124.6 & - \end{tabular} & $B_u^5$ & \begin{tabular}{>{\centering\arraybackslash}m{2.0cm} >{\centering\arraybackslash}m{2.0cm}} 160.4 & - \end{tabular} \\
$B_g^4$ & \begin{tabular}{>{\centering\arraybackslash}m{2.0cm} >{\centering\arraybackslash}m{2.0cm}} 127.2 & 130.4 \end{tabular} & $B_u^6$ & \begin{tabular}{>{\centering\arraybackslash}m{2.0cm} >{\centering\arraybackslash}m{2.0cm}} 188.6 & - \end{tabular} \\
$A_g^6$ & \begin{tabular}{>{\centering\arraybackslash}m{2.0cm} >{\centering\arraybackslash}m{2.0cm}} 149.8 & 155 \end{tabular} & $B_u^7$ & \begin{tabular}{>{\centering\arraybackslash}m{2.0cm} >{\centering\arraybackslash}m{2.0cm}} 227.9 & - \end{tabular} \\
$A_g^7$ & \begin{tabular}{>{\centering\arraybackslash}m{2.0cm} >{\centering\arraybackslash}m{2.0cm}} 164.9 & 165 \end{tabular} & $A_u^4$ & \begin{tabular}{>{\centering\arraybackslash}m{2.0cm} >{\centering\arraybackslash}m{2.0cm}} 231.1 & - \end{tabular} \\
$A_g^8$ & \begin{tabular}{>{\centering\arraybackslash}m{2.0cm} >{\centering\arraybackslash}m{2.0cm}} 191 & - \end{tabular} & $B_u^8$ & \begin{tabular}{>{\centering\arraybackslash}m{2.0cm} >{\centering\arraybackslash}m{2.0cm}} 289.4 & - \end{tabular} \\
$B_g^5$ & \begin{tabular}{>{\centering\arraybackslash}m{2.0cm} >{\centering\arraybackslash}m{2.0cm}} 217.1 & 222.3 \end{tabular} & & \\
$A_g^9$ & \begin{tabular}{>{\centering\arraybackslash}m{2.0cm} >{\centering\arraybackslash}m{2.0cm}} 241.9 & 223.9 \end{tabular} & & \\
$A_g^{10}$ & \begin{tabular}{>{\centering\arraybackslash}m{2.0cm} >{\centering\arraybackslash}m{2.0cm}} 276.22 & - \end{tabular} & & \\

\end{tabular}}
\end{ruledtabular}
\end{table*}

TaFe$_{1+y}$Te$_3$ crystallizes in the monoclinic crystal structure, which is shown in Fig.~\ref{fig1b}. The space group is $P2_1/m$ (unique axis $b$), with two formula units per unit cell.\cite{Badding_JSCS1992, Neuhausen_ZNB1993} The experimental values of the unit cell parameters are $a=7.436$ \AA, $b=3.638$ \AA, $c=10.008$ \AA, $\beta=109.17^{\circ}$. All atoms (including the excess Fe), are at $2e$ Wyckoff positions, with fractional coordinates given in Table~\ref{tab1}.

The factor group analysis (FGA) for $P2_1/m$ space group yields a following normal mode distribution at the $\Gamma$-point:
\begin{eqnarray*}
\Gamma_{\text{Raman}}=10A_g+5B_g \\
\Gamma_{\text{IR}}=4A_u+8B_u \\
\Gamma_{\text{acoustic}}=A_u+2B_u 
\end{eqnarray*} 
The Raman spectra were measured from the $(\bar{1}01)$-plane of the sample, which is the natural cleavage plane.\cite{Ke_PRB2012, Min_CPL2015} From the Raman tensors given in Table~\ref{tab2}, the $A_g$ phonon modes are expected to be observable in the $(xx)$ and $(yy)$ scattering configurations. The $B_g$ modes can be observed only in $(xy)$ polarization configuration.

Raman scattering spectra of TaFe$_{1.25}$Te$_3$ single crystals, measured at 100 K in three different polarization configurations, are presented in Figure~\ref{fig2}. By using the selection rules, we assign the Raman peaks appearing in the $(xx)$ and $(yy)$ polarization configuration as the $A_g$ ones. This conclusion is supported by the lattice dynamics calculations, given in Table~\ref{tab2}. By comparing the calculated values of $A_g$ mode energies with those of the peaks appearing in the $(xx)$ and $(yy)$ spectra, we can unambiguously assign four Raman modes ($A_g^4, A_g^6, A_g^7$ and $A_g^9$). The broad structure around 65 cm$^{-1}$ probably originates from the $A_g^2$ and $A_g^3$ modes, although the contribution of the $A_g^1$ mode (with calculated energy of 42.7 cm$^{-1}$) cannot be excluded. The peaks at 57.9 cm$^{-1}$ and 130 cm$^{-1}$ that are clearly visible in $(xy)$ but absent in $(yy)$ configuration are assigned as $B_g^2$ and $B_g^4$ modes, respectively. The low intensity peak at $\approx$ 220 cm$^{-1}$, that becomes clearly observable at low temperatures, is tentatively assigned as $B_g^5$ mode, although the contribution from the leakage of $A_g^9$ mode cannot be excluded. The origin of the two very broad structures at about 70 cm$^{-1}$ and 160 cm$^{-1}$, which are pronounced in the $(xy)$ configuration, is not completely clear. Aside to providing additional charge, Fe2 atoms may contribute to momentum transfer scattering, in line with the pronounced quasi-elastic continuum, present in all the scattering configurations. Consequently, contribution from single-phonon scattering away from $\Gamma$ point becomes observable, which is theoretically predicted\cite{Shuker_PRL1970, Benassi_PRB1991} and experimentally observed.\cite{Ryu2_PRB2015, Zhang_AIP2015} Although, we can not exclude the possibility of two- and, in particular, double-phonon contributions, we believe it is less likely due to the nature of the processes and since they usually have more pronounced contribution to $A$ channel (for arbitrary irreducible symmetry $\mu$ of $C_{2h}$ holds $\mu\otimes\mu\ni A$).

\begin{figure}
\centering
\includegraphics[width = 85mm]{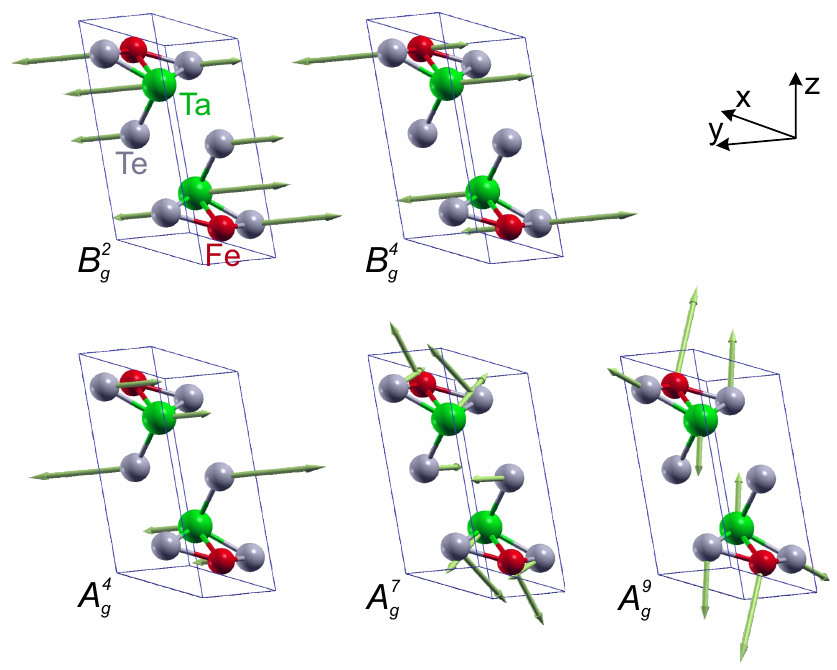}
\caption{Unit cell of TaFeTe$_3$ single crystal with the displacement patterns of several $A_g$ and $B_g$ Raman modes. Arrow lengths are proportional to the square root of the interatomic forces.}
\label{fig3}
\end{figure}

The normal modes of the selected $A_g$ and $B_g$ vibrations, as obtained by the lattice dynamics calculations, are shown in Fig.~\ref{fig3}. The low energy $B_g^2$ mode represents vibrations of Te and Ta atoms which tend to elongate the (Ta,Fe)Te tetrahedra in the $xy$-plane. $B_g^4$ phonon originates mainly from Ta and Te atom vibrations in directions opposite to each other, whereas $A_g^4$ mode represents dominantly vibrations of another Te atom in the $xy$-plane. $A_g^7$ and $A_g^9$ modes originate from the vibrations of Fe and Te atoms which tend to rotate the tetrahedra around the $x$-axis.

The DFT calculations are in very good agreement with the measured Raman spectra, specially having in mind the strength of electronic correlations in iron based compounds and the presence of additional Fe2 atoms in the measured sample. We restricted to the non-magnetic DFT calculations. This is because small changes in the phonon energies due to the magnetic ordering cannot be reliably captured since the DFT does not treat strong correlation and spin-fluctuations effects. Our DFT calculations for the electronic band structure agree with the results from Ref.~\onlinecite{Min_CPL2015}. The calculated electronic dispersions are in rather good agreement with the ARPES measurements,\cite{Min_CPL2015} which indicates that the main effect of the interstitial Fe2 ion is to provide additional charge and shift the Fermi level. This conclusion is supported with a small difference between the relaxed and experimental fractional coordinates, see Table~\ref{tab1}.

\begin{figure}[t]
\centering
\includegraphics[width = 85mm]{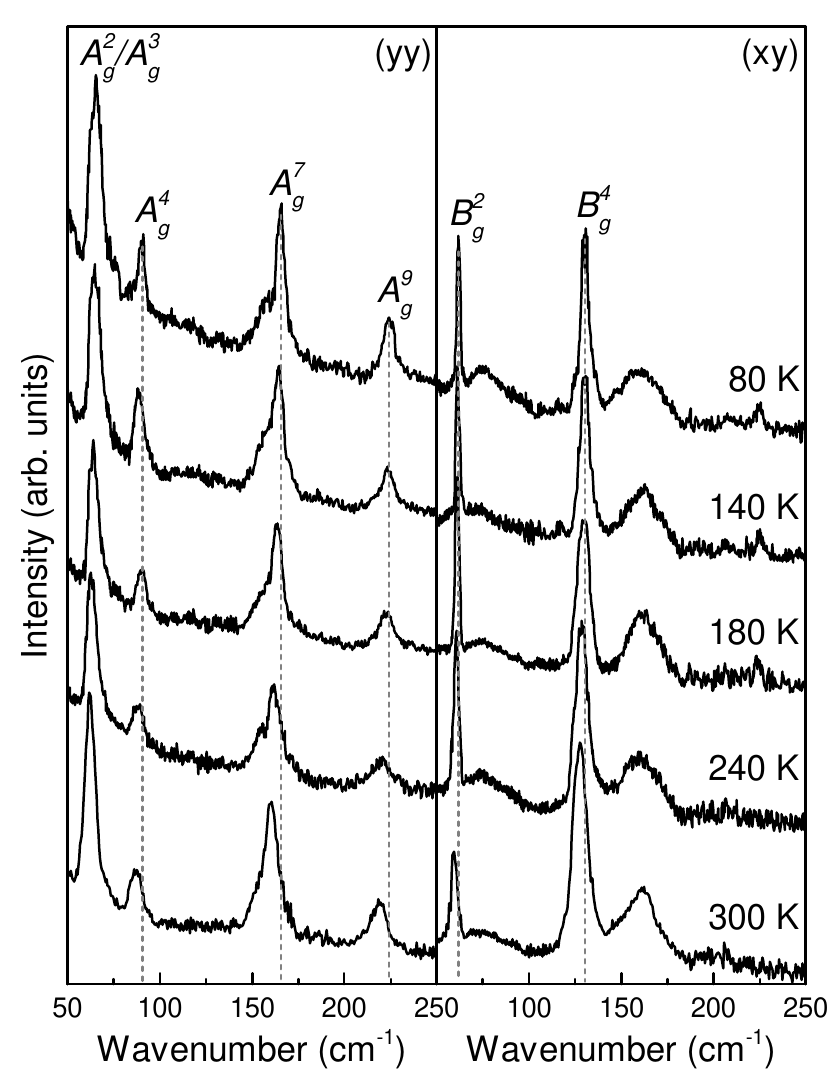}
\caption{(Color online) Temperature dependent Raman scattering spectra of TaFe$_{1.25}$Te$_3$ single crystal in the $(yy)$ (left panel) and $(xy)$ (right panel) polarization.}
\label{fig4}
\end{figure}

In order to analyze the changes of the Raman spectra near the AFM transition at $T_N\approx 200$ K, we have performed measurements in a temperature range from 80 K up to 300 K. Raman spectra of TaFe$_{1.25}$Te$_3$ single crystal, measured at different temperatures in the $(yy)$ and $(xy)$ scattering configurations, are given in Figure~\ref{fig4}. In the following, we perform the temperature analysis of the energy and the linewidth for five most clearly observed modes.

\begin{figure}
\centering
\includegraphics[width = 85mm]{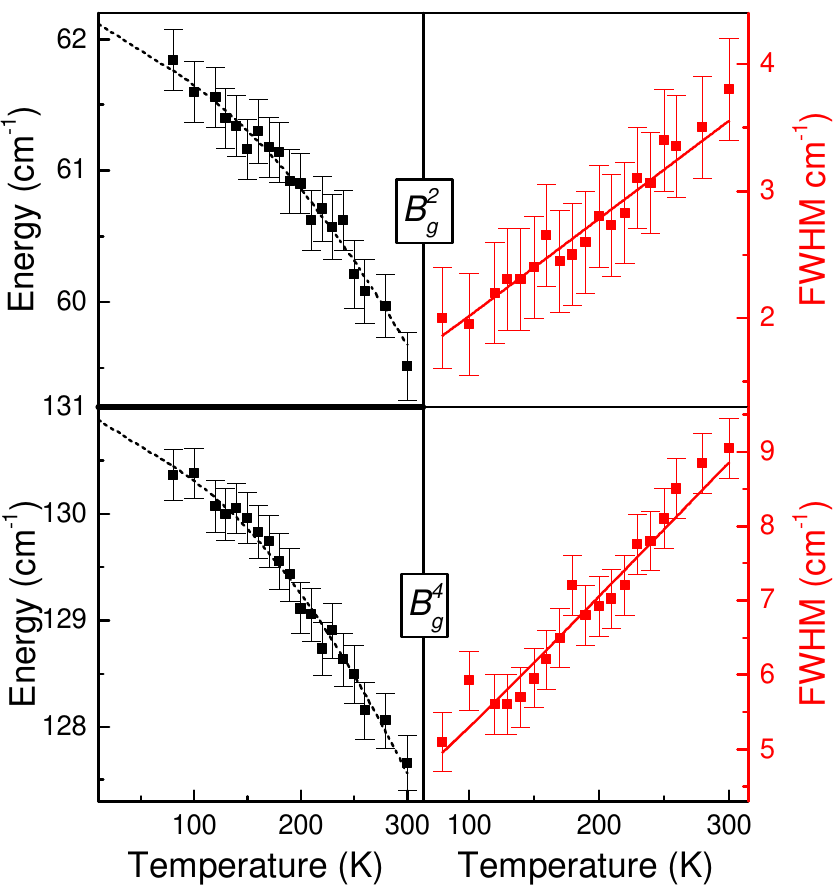}
\caption{(Color online) Temperature dependence of the energy and linewidth for the $B_g^2$ and $B_g^4$ Raman modes of the TaFe$_{1.25}$Te$_3$ single crystal. The red lines are fitted according to Eq.~(\ref{eq4}), whereas black lines are guide to the eye.}
\label{fig5}
\end{figure}

\begin{figure}[t]
\centering
\includegraphics[width = 85mm]{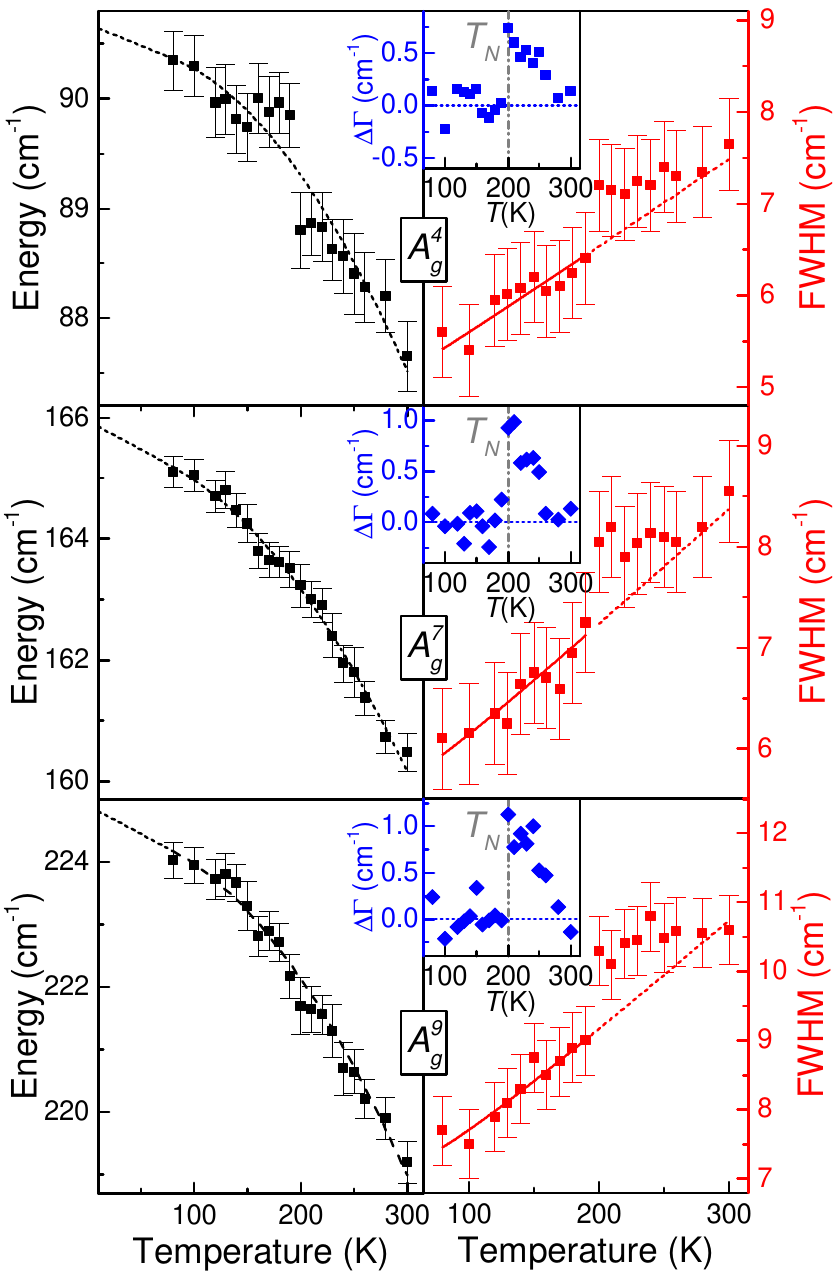}
\caption{(Color online) Energy and linewidth of the $A_g^4, A_g^7$ and $A_g^9$ Raman modes of the TaFe$_{1.25}$Te$_3$ single crystal as a function of temperature. The red lines are plotted according to Eq.~(\ref{eq4}), and the black dotted lines are guide to the eye. The insets represent deviations of the Raman mode linewidth from the anharmonic form.}
\label{fig6}
\end{figure}

The temperature dependence of the Raman mode energy is usually described with\cite{Menendez_PRB1984,Eiter_PRB2014}
\begin{equation}
\omega_i(T)=\omega_{0,i}+\Delta_i^V(\gamma_i, \alpha_i(T))+\Delta_i^A(C_i),
\label{eq1}
\end{equation}
\noindent where $\omega_{0,i}$ is a temperature independent contribution to the Raman mode energy. The second term represents a change of the phonon energy induced by the lattice thermal expansion and depends on the Gr\"{u}neisen parameter $\gamma_i$ and the thermal expansion coefficient $\alpha_i(T)$. The term $\Delta_i^A$ describes the anharmonicity induced change of the Raman mode energy which is a function of the anharmonic constant $C_i$. Both $\Delta_i^V$ and $\Delta_i^A$ have qualitatively the same temperature dependence. Since there are no reported experimental data on the temperature dependence of the lattice parameters for TaFe$_{1+y}$Te$_3$, we didn't attempt to fit the data, and the black dotted lines in Figures~\ref{fig5} and~\ref{fig6} are guides to the eye. The $\omega_i(T)$ curves follow the ''standard''\cite{Litvinchuk_PRB2011,Rahlenbeck_PRB2009,Lazarevic_PRB2013,Opacic_JPCM2015,Opacic_JPCM2016} continuous decrease in energy with temperature, with very small anomalies near $T_N$ except for the A$_g^4$ mode.

The temperature dependences of the linewidth of selected $B_g$ and $A_g$ modes are given in the right panels of Figs.~\ref{fig5} and~\ref{fig6}, respectively. While the $B_g^2$ and $B_g^4$ phonon modes do not show significant deviation from the usual behavior due to the anharmonicity effects, with gradual broadening with increasing temperature, the $A_g^4, A_g^7$ and $A_g^9$ modes exhibit moderate additional broadening above 200 K. The red lines present a fit to the standard formula for the temperature dependent linewidth due to the anharmonicity:\cite{Menendez_PRB1984, Gnezdilov_PRB2011, Iliev_PRB1999}
\begin{equation}
\Gamma_i(T)=\Gamma_{0,i}\left(1+\frac{2}{e^{\hbar \omega_{0,i}/2k_BT}-1}\right) + A_i,
\label{eq4}
\end{equation}
where $\Gamma_{0,i}$ is the anharmonic constant and $A_i$ is the constant term due to the disorder and electron-phonon interaction.\cite{Lazarevic:2010}
The deviation from these anharmonicity curves is most pronounced around $T_N$ (see the insets of Fig.~\ref{fig6}).

We can observe that all Raman modes have moderate linewidth and exhibit small anomalies near $T_N$. This shows that the phase transition is continuous, in agreement with the thermodynamic and transport measurements.\cite{Liu_PRB2011} Small anomalies in the phonon spectra, which are restricted only to the vicinity of the phase transition, imply that the electron-phonon interaction of Raman active modes does not change with temperature. This is in agreement with the recent ARPES measurements which show negligible change of the Fermi surface across the AFM transition,\cite{Min_CPL2015} indicating that the magnetic transition is not driven by the Fermi surface instability. The anomalies in the linewidth of some phonon modes near $T_N$ are likely the signature of the increased scattering by spin fluctuations near the phase transition.\cite{Iliev_PRB1999, Litvinchuk_PRB1999}

The density of states (DOS) at the Fermi level is not large. This can be concluded from the ARPES experiments\cite{Min_CPL2015} which have shown three bands crossing the Fermi level but with strong dispersion, while several relatively flat bands are found only well below the Fermi level. The DFT calculations also give moderate values for the DOS, $N(E_F) \approx 1$  eV$^{-1}$/f.u., after the Fermi level is shifted due to the additional charge provided by the Fe2 atoms. This value for the DOS also suggests that the electron-phonon coupling is not strong in TaFe$_{1.25}$Te$_3$, since it is proportional to $N(E_F)$.

TaFe$_{1.25}$Te$_3$ has a similar moment size as Fe$_{1+y}$Te, $\approx 2 \mu_B$/Fe. However, the differences in the magnetic ordering and crystal structure cause different phonon properties of these two compounds. Namely, the phonon lines in the Raman spectra of Fe$_{1+y}$Te have very large linewidth and pronounced anomalies both in the frequency and in the linewidth near the first order phase transition.\cite{Gnezdilov_PRB2011, Popovic_SSC2014} Small anomalies in the Raman spectra of  TaFe$_{1.25}$Te$_3$ as compared to  Fe$_{1+y}$Te can be ascribed to the continuous, second order nature of the AFM transition and smaller electron-phonon coupling due to lower DOS at the Fermi level. Also, the monoclinic angle $\beta$ in the TaFe$_{1.25}$Te$_3$ unit cell significantly differs from $90^\circ$ and therefore the form of the vibrational modes is different.

\section{Conclusion}

In summary,we have performed the Raman scattering study of the zigzag spin chain TaFe$_{1.25}$Te$_3$ single crystal, together with the lattice dynamics calculations of TaFeTe$_3$. By analyzing the Raman spectra in different polarization configurations and using numerical calculations we have assigned 9 Raman active modes predicted by the FGA.  Very good agreement between the experimental frequencies and those calculated for the stoichiometric compound shows that the excess iron atoms weakly influence the phonon energies but provide momentum conservation for the phonon scattering away from $\Gamma$ point. The temperature dependence of the frequency and the linewidth of the $B_g$ Raman modes looks conventional, governed by the anharmonicity effects. While in a broad temperature range the behavior of the $A_g$ modes is also conventional, there are clear anomalies near the AFM transition. The anomalies in the frequency and the linewidth are in the form of small kinks near $T_N$. This implies that the electron-phonon interaction and the DOS at the Fermi level are approximately constant in the measured temperature range. The increase in the linewidth near $T_N$ is likely due to the coupling of spin fluctuations and vibration near the second order phase transition.

\begin{acknowledgments}
We thank D. Stepanenko for useful discussions.
This work was supported by the Serbian Ministry of Education, Science and Technological Development under Projects ON171032, III45018 and ON171017.  Work at Brookhaven is supported by the Center for Emergent Superconductivity, an Energy Frontier Research Center funded by the U.S. DOE, Office for Basic Energy Science (C.P.). Numerical calculations were performed on the PARADOX supercomputing facility at the Scientific Computing Laboratory of the Institute of Physics Belgrade.
\end{acknowledgments}


\begin{thebibliography}{53}%
\makeatletter
\providecommand \@ifxundefined [1]{%
 \@ifx{#1\undefined}
}%
\providecommand \@ifnum [1]{%
 \ifnum #1\expandafter \@firstoftwo
 \else \expandafter \@secondoftwo
 \fi
}%
\providecommand \@ifx [1]{%
 \ifx #1\expandafter \@firstoftwo
 \else \expandafter \@secondoftwo
 \fi
}%
\providecommand \natexlab [1]{#1}%
\providecommand \enquote  [1]{``#1''}%
\providecommand \bibnamefont  [1]{#1}%
\providecommand \bibfnamefont [1]{#1}%
\providecommand \citenamefont [1]{#1}%
\providecommand \href@noop [0]{\@secondoftwo}%
\providecommand \href [0]{\begingroup \@sanitize@url \@href}%
\providecommand \@href[1]{\@@startlink{#1}\@@href}%
\providecommand \@@href[1]{\endgroup#1\@@endlink}%
\providecommand \@sanitize@url [0]{\catcode `\\12\catcode `\$12\catcode
  `\&12\catcode `\#12\catcode `\^12\catcode `\_12\catcode `\%12\relax}%
\providecommand \@@startlink[1]{}%
\providecommand \@@endlink[0]{}%
\providecommand \url  [0]{\begingroup\@sanitize@url \@url }%
\providecommand \@url [1]{\endgroup\@href {#1}{\urlprefix }}%
\providecommand \urlprefix  [0]{URL }%
\providecommand \Eprint [0]{\href }%
\providecommand \doibase [0]{http://dx.doi.org/}%
\providecommand \selectlanguage [0]{\@gobble}%
\providecommand \bibinfo  [0]{\@secondoftwo}%
\providecommand \bibfield  [0]{\@secondoftwo}%
\providecommand \translation [1]{[#1]}%
\providecommand \BibitemOpen [0]{}%
\providecommand \bibitemStop [0]{}%
\providecommand \bibitemNoStop [0]{.\EOS\space}%
\providecommand \EOS [0]{\spacefactor3000\relax}%
\providecommand \BibitemShut  [1]{\csname bibitem#1\endcsname}%
\let\auto@bib@innerbib\@empty
\bibitem [{\citenamefont {Kamihara}\ \emph {et~al.}(2008)\citenamefont
  {Kamihara}, \citenamefont {Watanabe}, \citenamefont {Hirano},\ and\
  \citenamefont {Hosono}}]{Kamihara_JACS2008}%
  \BibitemOpen
  \bibfield  {author} {\bibinfo {author} {\bibfnamefont {Y.}~\bibnamefont
  {Kamihara}}, \bibinfo {author} {\bibfnamefont {T.}~\bibnamefont {Watanabe}},
  \bibinfo {author} {\bibfnamefont {M.}~\bibnamefont {Hirano}}, \ and\ \bibinfo
  {author} {\bibfnamefont {H.}~\bibnamefont {Hosono}},\ }\href {\doibase
  10.1021/ja800073m} {\bibfield  {journal} {\bibinfo  {journal} {Journal of the
  American Chemical Society}\ }\textbf {\bibinfo {volume} {130}},\ \bibinfo
  {pages} {3296} (\bibinfo {year} {2008})}\BibitemShut {NoStop}%
\bibitem [{\citenamefont {Ma}\ \emph {et~al.}(2012)\citenamefont {Ma},
  \citenamefont {Ji}, \citenamefont {Dai}, \citenamefont {Lu}, \citenamefont
  {Eom}, \citenamefont {Kim}, \citenamefont {Normand},\ and\ \citenamefont
  {Yu}}]{Ma_PRL2012}%
  \BibitemOpen
  \bibfield  {author} {\bibinfo {author} {\bibfnamefont {L.}~\bibnamefont
  {Ma}}, \bibinfo {author} {\bibfnamefont {G.~F.}\ \bibnamefont {Ji}}, \bibinfo
  {author} {\bibfnamefont {J.}~\bibnamefont {Dai}}, \bibinfo {author}
  {\bibfnamefont {X.~R.}\ \bibnamefont {Lu}}, \bibinfo {author} {\bibfnamefont
  {M.~J.}\ \bibnamefont {Eom}}, \bibinfo {author} {\bibfnamefont {J.~S.}\
  \bibnamefont {Kim}}, \bibinfo {author} {\bibfnamefont {B.}~\bibnamefont
  {Normand}}, \ and\ \bibinfo {author} {\bibfnamefont {W.}~\bibnamefont {Yu}},\
  }\href {\doibase 10.1103/PhysRevLett.109.197002} {\bibfield  {journal}
  {\bibinfo  {journal} {Phys. Rev. Lett.}\ }\textbf {\bibinfo {volume} {109}},\
  \bibinfo {pages} {197002} (\bibinfo {year} {2012})}\BibitemShut {NoStop}%
\bibitem [{\citenamefont {Wei}\ \emph {et~al.}(2011)\citenamefont {Wei},
  \citenamefont {Qing-Zhen}, \citenamefont {Gen-Fu}, \citenamefont {Green},
  \citenamefont {Du-Ming}, \citenamefont {Jun-Bao},\ and\ \citenamefont
  {Yi-Ming}}]{Bao_ChinPhysLett2011}%
  \BibitemOpen
  \bibfield  {author} {\bibinfo {author} {\bibfnamefont {B.}~\bibnamefont
  {Wei}}, \bibinfo {author} {\bibfnamefont {H.}~\bibnamefont {Qing-Zhen}},
  \bibinfo {author} {\bibfnamefont {C.}~\bibnamefont {Gen-Fu}}, \bibinfo
  {author} {\bibfnamefont {M.~A.}\ \bibnamefont {Green}}, \bibinfo {author}
  {\bibfnamefont {W.}~\bibnamefont {Du-Ming}}, \bibinfo {author} {\bibfnamefont
  {H.}~\bibnamefont {Jun-Bao}}, \ and\ \bibinfo {author} {\bibfnamefont
  {Q.}~\bibnamefont {Yi-Ming}},\ }\href
  {http://stacks.iop.org/0256-307X/28/i=8/a=086104} {\bibfield  {journal}
  {\bibinfo  {journal} {Chinese Physics Letters}\ }\textbf {\bibinfo {volume}
  {28}},\ \bibinfo {pages} {086104} (\bibinfo {year} {2011})}\BibitemShut
  {NoStop}%
\bibitem [{\citenamefont {Stewart}(2011)}]{Stewart_RevModPhys2011}%
  \BibitemOpen
  \bibfield  {author} {\bibinfo {author} {\bibfnamefont {G.~R.}\ \bibnamefont
  {Stewart}},\ }\href {\doibase 10.1103/RevModPhys.83.1589} {\bibfield
  {journal} {\bibinfo  {journal} {Rev. Mod. Phys.}\ }\textbf {\bibinfo {volume}
  {83}},\ \bibinfo {pages} {1589} (\bibinfo {year} {2011})}\BibitemShut
  {NoStop}%
\bibitem [{\citenamefont {Opa\v{c}i\'c}\ \emph {et~al.}(2016)\citenamefont
  {Opa\v{c}i\'c}, \citenamefont {Lazarevi\'c}, \citenamefont {Radonji\'c},
  \citenamefont {\v{S}\'cepanovi\'c}, \citenamefont {Ryu}, \citenamefont
  {Wang}, \citenamefont {Tanaskovi\'c}, \citenamefont {Petrovic},\ and\
  \citenamefont {Popovi\'c}}]{Opacic_JPCM2016}%
  \BibitemOpen
  \bibfield  {author} {\bibinfo {author} {\bibfnamefont {M.}~\bibnamefont
  {Opa\v{c}i\'c}}, \bibinfo {author} {\bibfnamefont {N.}~\bibnamefont
  {Lazarevi\'c}}, \bibinfo {author} {\bibfnamefont {M.~M.}\ \bibnamefont
  {Radonji\'c}}, \bibinfo {author} {\bibfnamefont {M.}~\bibnamefont
  {\v{S}\'cepanovi\'c}}, \bibinfo {author} {\bibfnamefont {H.}~\bibnamefont
  {Ryu}}, \bibinfo {author} {\bibfnamefont {A.}~\bibnamefont {Wang}}, \bibinfo
  {author} {\bibfnamefont {D.}~\bibnamefont {Tanaskovi\'c}}, \bibinfo {author}
  {\bibfnamefont {C.}~\bibnamefont {Petrovic}}, \ and\ \bibinfo {author}
  {\bibfnamefont {Z.~V.}\ \bibnamefont {Popovi\'c}},\ }\href
  {http://stacks.iop.org/0953-8984/28/i=48/a=485401} {\bibfield  {journal}
  {\bibinfo  {journal} {Journal of Physics: Condensed Matter}\ }\textbf
  {\bibinfo {volume} {28}},\ \bibinfo {pages} {485401} (\bibinfo {year}
  {2016})}\BibitemShut {NoStop}%
\bibitem [{\citenamefont {G\"{o}nen}\ \emph {et~al.}(2000)\citenamefont
  {G\"{o}nen}, \citenamefont {Fournier}, \citenamefont {Smolyaninova},
  \citenamefont {Greene}, \citenamefont {Araujo-Moreira},\ and\ \citenamefont
  {Eichhorn}}]{Gonen_CHM2000}%
  \BibitemOpen
  \bibfield  {author} {\bibinfo {author} {\bibfnamefont {Z.~S.}\ \bibnamefont
  {G\"{o}nen}}, \bibinfo {author} {\bibfnamefont {P.}~\bibnamefont {Fournier}},
  \bibinfo {author} {\bibfnamefont {V.}~\bibnamefont {Smolyaninova}}, \bibinfo
  {author} {\bibfnamefont {R.}~\bibnamefont {Greene}}, \bibinfo {author}
  {\bibfnamefont {F.~M.}\ \bibnamefont {Araujo-Moreira}}, \ and\ \bibinfo
  {author} {\bibfnamefont {B.}~\bibnamefont {Eichhorn}},\ }\href {\doibase
  10.1021/cm0000346} {\bibfield  {journal} {\bibinfo  {journal} {Chemistry of
  Materials}\ }\textbf {\bibinfo {volume} {12}},\ \bibinfo {pages} {3331}
  (\bibinfo {year} {2000})}\BibitemShut {NoStop}%
\bibitem [{\citenamefont {Lei}\ \emph {et~al.}(2011)\citenamefont {Lei},
  \citenamefont {Ryu}, \citenamefont {Frenkel},\ and\ \citenamefont
  {Petrovic}}]{Lei_PRB2011}%
  \BibitemOpen
  \bibfield  {author} {\bibinfo {author} {\bibfnamefont {H.}~\bibnamefont
  {Lei}}, \bibinfo {author} {\bibfnamefont {H.}~\bibnamefont {Ryu}}, \bibinfo
  {author} {\bibfnamefont {A.~I.}\ \bibnamefont {Frenkel}}, \ and\ \bibinfo
  {author} {\bibfnamefont {C.}~\bibnamefont {Petrovic}},\ }\href {\doibase
  10.1103/PhysRevB.84.214511} {\bibfield  {journal} {\bibinfo  {journal} {Phys.
  Rev. B}\ }\textbf {\bibinfo {volume} {84}},\ \bibinfo {pages} {214511}
  (\bibinfo {year} {2011})}\BibitemShut {NoStop}%
\bibitem [{\citenamefont {Han}\ \emph {et~al.}(2012)\citenamefont {Han},
  \citenamefont {Wan}, \citenamefont {Shen},\ and\ \citenamefont
  {Wen}}]{Han_PRB2012}%
  \BibitemOpen
  \bibfield  {author} {\bibinfo {author} {\bibfnamefont {F.}~\bibnamefont
  {Han}}, \bibinfo {author} {\bibfnamefont {X.}~\bibnamefont {Wan}}, \bibinfo
  {author} {\bibfnamefont {B.}~\bibnamefont {Shen}}, \ and\ \bibinfo {author}
  {\bibfnamefont {H.-H.}\ \bibnamefont {Wen}},\ }\href {\doibase
  10.1103/PhysRevB.86.014411} {\bibfield  {journal} {\bibinfo  {journal} {Phys.
  Rev. B}\ }\textbf {\bibinfo {volume} {86}},\ \bibinfo {pages} {014411}
  (\bibinfo {year} {2012})}\BibitemShut {NoStop}%
\bibitem [{\citenamefont {Fang}\ \emph {et~al.}(2008)\citenamefont {Fang},
  \citenamefont {Pham}, \citenamefont {Qian}, \citenamefont {Liu},
  \citenamefont {Vehstedt}, \citenamefont {Liu}, \citenamefont {Spinu},\ and\
  \citenamefont {Mao}}]{Fang_PRB2008}%
  \BibitemOpen
  \bibfield  {author} {\bibinfo {author} {\bibfnamefont {M.~H.}\ \bibnamefont
  {Fang}}, \bibinfo {author} {\bibfnamefont {H.~M.}\ \bibnamefont {Pham}},
  \bibinfo {author} {\bibfnamefont {B.}~\bibnamefont {Qian}}, \bibinfo {author}
  {\bibfnamefont {T.~J.}\ \bibnamefont {Liu}}, \bibinfo {author} {\bibfnamefont
  {E.~K.}\ \bibnamefont {Vehstedt}}, \bibinfo {author} {\bibfnamefont
  {Y.}~\bibnamefont {Liu}}, \bibinfo {author} {\bibfnamefont {L.}~\bibnamefont
  {Spinu}}, \ and\ \bibinfo {author} {\bibfnamefont {Z.~Q.}\ \bibnamefont
  {Mao}},\ }\href {\doibase 10.1103/PhysRevB.78.224503} {\bibfield  {journal}
  {\bibinfo  {journal} {Phys. Rev. B}\ }\textbf {\bibinfo {volume} {78}},\
  \bibinfo {pages} {224503} (\bibinfo {year} {2008})}\BibitemShut {NoStop}%
\bibitem [{\citenamefont {Li}\ \emph {et~al.}(2009)\citenamefont {Li},
  \citenamefont {de~la Cruz}, \citenamefont {Huang}, \citenamefont {Chen},
  \citenamefont {Lynn}, \citenamefont {Hu}, \citenamefont {Huang},
  \citenamefont {Hsu}, \citenamefont {Yeh}, \citenamefont {Wu},\ and\
  \citenamefont {Dai}}]{Li_PRB2009}%
  \BibitemOpen
  \bibfield  {author} {\bibinfo {author} {\bibfnamefont {S.}~\bibnamefont
  {Li}}, \bibinfo {author} {\bibfnamefont {C.}~\bibnamefont {de~la Cruz}},
  \bibinfo {author} {\bibfnamefont {Q.}~\bibnamefont {Huang}}, \bibinfo
  {author} {\bibfnamefont {Y.}~\bibnamefont {Chen}}, \bibinfo {author}
  {\bibfnamefont {J.~W.}\ \bibnamefont {Lynn}}, \bibinfo {author}
  {\bibfnamefont {J.}~\bibnamefont {Hu}}, \bibinfo {author} {\bibfnamefont
  {Y.-L.}\ \bibnamefont {Huang}}, \bibinfo {author} {\bibfnamefont {F.-C.}\
  \bibnamefont {Hsu}}, \bibinfo {author} {\bibfnamefont {K.-W.}\ \bibnamefont
  {Yeh}}, \bibinfo {author} {\bibfnamefont {M.-K.}\ \bibnamefont {Wu}}, \ and\
  \bibinfo {author} {\bibfnamefont {P.}~\bibnamefont {Dai}},\ }\href {\doibase
  10.1103/PhysRevB.79.054503} {\bibfield  {journal} {\bibinfo  {journal} {Phys.
  Rev. B}\ }\textbf {\bibinfo {volume} {79}},\ \bibinfo {pages} {054503}
  (\bibinfo {year} {2009})}\BibitemShut {NoStop}%
\bibitem [{\citenamefont {Gnezdilov}\ \emph {et~al.}(2011)\citenamefont
  {Gnezdilov}, \citenamefont {Pashkevich}, \citenamefont {Lemmens},
  \citenamefont {Gusev}, \citenamefont {Lamonova}, \citenamefont {Shevtsova},
  \citenamefont {Vitebskiy}, \citenamefont {Afanasiev}, \citenamefont
  {Gnatchenko}, \citenamefont {Tsurkan}, \citenamefont {Deisenhofer},\ and\
  \citenamefont {Loidl}}]{Gnezdilov_PRB2011}%
  \BibitemOpen
  \bibfield  {author} {\bibinfo {author} {\bibfnamefont {V.}~\bibnamefont
  {Gnezdilov}}, \bibinfo {author} {\bibfnamefont {Y.}~\bibnamefont
  {Pashkevich}}, \bibinfo {author} {\bibfnamefont {P.}~\bibnamefont {Lemmens}},
  \bibinfo {author} {\bibfnamefont {A.}~\bibnamefont {Gusev}}, \bibinfo
  {author} {\bibfnamefont {K.}~\bibnamefont {Lamonova}}, \bibinfo {author}
  {\bibfnamefont {T.}~\bibnamefont {Shevtsova}}, \bibinfo {author}
  {\bibfnamefont {I.}~\bibnamefont {Vitebskiy}}, \bibinfo {author}
  {\bibfnamefont {O.}~\bibnamefont {Afanasiev}}, \bibinfo {author}
  {\bibfnamefont {S.}~\bibnamefont {Gnatchenko}}, \bibinfo {author}
  {\bibfnamefont {V.}~\bibnamefont {Tsurkan}}, \bibinfo {author} {\bibfnamefont
  {J.}~\bibnamefont {Deisenhofer}}, \ and\ \bibinfo {author} {\bibfnamefont
  {A.}~\bibnamefont {Loidl}},\ }\href {\doibase 10.1103/PhysRevB.83.245127}
  {\bibfield  {journal} {\bibinfo  {journal} {Phys. Rev. B}\ }\textbf {\bibinfo
  {volume} {83}},\ \bibinfo {pages} {245127} (\bibinfo {year}
  {2011})}\BibitemShut {NoStop}%
\bibitem [{\citenamefont {Um}\ \emph {et~al.}(2012)\citenamefont {Um},
  \citenamefont {Subedi}, \citenamefont {Toulemonde}, \citenamefont {Ganin},
  \citenamefont {Boeri}, \citenamefont {Rahlenbeck}, \citenamefont {Liu},
  \citenamefont {Lin}, \citenamefont {Carlsson}, \citenamefont {Sulpice},
  \citenamefont {Rosseinsky}, \citenamefont {Keimer},\ and\ \citenamefont
  {Le~Tacon}}]{Um_PRB2012}%
  \BibitemOpen
  \bibfield  {author} {\bibinfo {author} {\bibfnamefont {Y.~J.}\ \bibnamefont
  {Um}}, \bibinfo {author} {\bibfnamefont {A.}~\bibnamefont {Subedi}}, \bibinfo
  {author} {\bibfnamefont {P.}~\bibnamefont {Toulemonde}}, \bibinfo {author}
  {\bibfnamefont {A.~Y.}\ \bibnamefont {Ganin}}, \bibinfo {author}
  {\bibfnamefont {L.}~\bibnamefont {Boeri}}, \bibinfo {author} {\bibfnamefont
  {M.}~\bibnamefont {Rahlenbeck}}, \bibinfo {author} {\bibfnamefont
  {Y.}~\bibnamefont {Liu}}, \bibinfo {author} {\bibfnamefont {C.~T.}\
  \bibnamefont {Lin}}, \bibinfo {author} {\bibfnamefont {S.~J.~E.}\
  \bibnamefont {Carlsson}}, \bibinfo {author} {\bibfnamefont {A.}~\bibnamefont
  {Sulpice}}, \bibinfo {author} {\bibfnamefont {M.~J.}\ \bibnamefont
  {Rosseinsky}}, \bibinfo {author} {\bibfnamefont {B.}~\bibnamefont {Keimer}},
  \ and\ \bibinfo {author} {\bibfnamefont {M.}~\bibnamefont {Le~Tacon}},\
  }\href {\doibase 10.1103/PhysRevB.85.064519} {\bibfield  {journal} {\bibinfo
  {journal} {Phys. Rev. B}\ }\textbf {\bibinfo {volume} {85}},\ \bibinfo
  {pages} {064519} (\bibinfo {year} {2012})}\BibitemShut {NoStop}%
\bibitem [{\citenamefont {Popovi\ifmmode~\acute{c}\else \'{c}\fi{}}\ \emph
  {et~al.}(2014)\citenamefont {Popovi\ifmmode~\acute{c}\else \'{c}\fi{}},
  \citenamefont {Lazarevi\ifmmode~\acute{c}\else \'{c}\fi{}}, \citenamefont
  {Bogdanovi\ifmmode~\acute{c}\else \'{c}\fi{}}, \citenamefont
  {Radonji\ifmmode~\acute{c}\else \'{c}\fi{}}, \citenamefont
  {Tanaskovi\ifmmode~\acute{c}\else \'{c}\fi{}}, \citenamefont {Hu},
  \citenamefont {Lei},\ and\ \citenamefont {Petrovic}}]{Popovic_SSC2014}%
  \BibitemOpen
  \bibfield  {author} {\bibinfo {author} {\bibfnamefont {Z.~V.}\ \bibnamefont
  {Popovi\ifmmode~\acute{c}\else \'{c}\fi{}}}, \bibinfo {author} {\bibfnamefont
  {N.}~\bibnamefont {Lazarevi\ifmmode~\acute{c}\else \'{c}\fi{}}}, \bibinfo
  {author} {\bibfnamefont {S.}~\bibnamefont {Bogdanovi\ifmmode~\acute{c}\else
  \'{c}\fi{}}}, \bibinfo {author} {\bibfnamefont {M.~M.}\ \bibnamefont
  {Radonji\ifmmode~\acute{c}\else \'{c}\fi{}}}, \bibinfo {author}
  {\bibfnamefont {D.}~\bibnamefont {Tanaskovi\ifmmode~\acute{c}\else
  \'{c}\fi{}}}, \bibinfo {author} {\bibfnamefont {R.}~\bibnamefont {Hu}},
  \bibinfo {author} {\bibfnamefont {H.}~\bibnamefont {Lei}}, \ and\ \bibinfo
  {author} {\bibfnamefont {C.}~\bibnamefont {Petrovic}},\ }\href {\doibase
  http://dx.doi.org/10.1016/j.ssc.2014.05.025} {\bibfield  {journal} {\bibinfo
  {journal} {Solid State Communications}\ }\textbf {\bibinfo {volume} {193}},\
  \bibinfo {pages} {51 } (\bibinfo {year} {2014})}\BibitemShut {NoStop}%
\bibitem [{\citenamefont {Choi}\ \emph {et~al.}(2008)\citenamefont {Choi},
  \citenamefont {Wulferding}, \citenamefont {Lemmens}, \citenamefont {Ni},
  \citenamefont {Bud'ko},\ and\ \citenamefont {Canfield}}]{Choi_PRB2008}%
  \BibitemOpen
  \bibfield  {author} {\bibinfo {author} {\bibfnamefont {K.-Y.}\ \bibnamefont
  {Choi}}, \bibinfo {author} {\bibfnamefont {D.}~\bibnamefont {Wulferding}},
  \bibinfo {author} {\bibfnamefont {P.}~\bibnamefont {Lemmens}}, \bibinfo
  {author} {\bibfnamefont {N.}~\bibnamefont {Ni}}, \bibinfo {author}
  {\bibfnamefont {S.~L.}\ \bibnamefont {Bud'ko}}, \ and\ \bibinfo {author}
  {\bibfnamefont {P.~C.}\ \bibnamefont {Canfield}},\ }\href {\doibase
  10.1103/PhysRevB.78.212503} {\bibfield  {journal} {\bibinfo  {journal} {Phys.
  Rev. B}\ }\textbf {\bibinfo {volume} {78}},\ \bibinfo {pages} {212503}
  (\bibinfo {year} {2008})}\BibitemShut {NoStop}%
\bibitem [{\citenamefont {Rahlenbeck}\ \emph {et~al.}(2009)\citenamefont
  {Rahlenbeck}, \citenamefont {Sun}, \citenamefont {Sun}, \citenamefont {Lin},
  \citenamefont {Keimer},\ and\ \citenamefont {Ulrich}}]{Rahlenbeck_PRB2009}%
  \BibitemOpen
  \bibfield  {author} {\bibinfo {author} {\bibfnamefont {M.}~\bibnamefont
  {Rahlenbeck}}, \bibinfo {author} {\bibfnamefont {G.~L.}\ \bibnamefont {Sun}},
  \bibinfo {author} {\bibfnamefont {D.~L.}\ \bibnamefont {Sun}}, \bibinfo
  {author} {\bibfnamefont {C.~T.}\ \bibnamefont {Lin}}, \bibinfo {author}
  {\bibfnamefont {B.}~\bibnamefont {Keimer}}, \ and\ \bibinfo {author}
  {\bibfnamefont {C.}~\bibnamefont {Ulrich}},\ }\href {\doibase
  10.1103/PhysRevB.80.064509} {\bibfield  {journal} {\bibinfo  {journal} {Phys.
  Rev. B}\ }\textbf {\bibinfo {volume} {80}},\ \bibinfo {pages} {064509}
  (\bibinfo {year} {2009})}\BibitemShut {NoStop}%
\bibitem [{\citenamefont {Chen}\ \emph {et~al.}(2011)\citenamefont {Chen},
  \citenamefont {Xu}, \citenamefont {Ge}, \citenamefont {Zhang}, \citenamefont
  {Ye}, \citenamefont {Yang}, \citenamefont {Jiang}, \citenamefont {Xie},
  \citenamefont {Che}, \citenamefont {Zhang}, \citenamefont {Wang},
  \citenamefont {Chen}, \citenamefont {Shen}, \citenamefont {Hu},\ and\
  \citenamefont {Feng}}]{Chen_PRX2011}%
  \BibitemOpen
  \bibfield  {author} {\bibinfo {author} {\bibfnamefont {F.}~\bibnamefont
  {Chen}}, \bibinfo {author} {\bibfnamefont {M.}~\bibnamefont {Xu}}, \bibinfo
  {author} {\bibfnamefont {Q.~Q.}\ \bibnamefont {Ge}}, \bibinfo {author}
  {\bibfnamefont {Y.}~\bibnamefont {Zhang}}, \bibinfo {author} {\bibfnamefont
  {Z.~R.}\ \bibnamefont {Ye}}, \bibinfo {author} {\bibfnamefont {L.~X.}\
  \bibnamefont {Yang}}, \bibinfo {author} {\bibfnamefont {J.}~\bibnamefont
  {Jiang}}, \bibinfo {author} {\bibfnamefont {B.~P.}\ \bibnamefont {Xie}},
  \bibinfo {author} {\bibfnamefont {R.~C.}\ \bibnamefont {Che}}, \bibinfo
  {author} {\bibfnamefont {M.}~\bibnamefont {Zhang}}, \bibinfo {author}
  {\bibfnamefont {A.~F.}\ \bibnamefont {Wang}}, \bibinfo {author}
  {\bibfnamefont {X.~H.}\ \bibnamefont {Chen}}, \bibinfo {author}
  {\bibfnamefont {D.~W.}\ \bibnamefont {Shen}}, \bibinfo {author}
  {\bibfnamefont {J.~P.}\ \bibnamefont {Hu}}, \ and\ \bibinfo {author}
  {\bibfnamefont {D.~L.}\ \bibnamefont {Feng}},\ }\href {\doibase
  10.1103/PhysRevX.1.021020} {\bibfield  {journal} {\bibinfo  {journal} {Phys.
  Rev. X}\ }\textbf {\bibinfo {volume} {1}},\ \bibinfo {pages} {021020}
  (\bibinfo {year} {2011})}\BibitemShut {NoStop}%
\bibitem [{\citenamefont {Li}\ \emph {et~al.}(2012)\citenamefont {Li},
  \citenamefont {Ding}, \citenamefont {Deng}, \citenamefont {Chang},
  \citenamefont {Song}, \citenamefont {He}, \citenamefont {Wang}, \citenamefont
  {Ma}, \citenamefont {Hu}, \citenamefont {Chen},\ and\ \citenamefont
  {Xue}}]{Li_NatPhys2012}%
  \BibitemOpen
  \bibfield  {author} {\bibinfo {author} {\bibfnamefont {W.}~\bibnamefont
  {Li}}, \bibinfo {author} {\bibfnamefont {H.}~\bibnamefont {Ding}}, \bibinfo
  {author} {\bibfnamefont {P.}~\bibnamefont {Deng}}, \bibinfo {author}
  {\bibfnamefont {K.}~\bibnamefont {Chang}}, \bibinfo {author} {\bibfnamefont
  {C.}~\bibnamefont {Song}}, \bibinfo {author} {\bibfnamefont {K.}~\bibnamefont
  {He}}, \bibinfo {author} {\bibfnamefont {L.}~\bibnamefont {Wang}}, \bibinfo
  {author} {\bibfnamefont {X.}~\bibnamefont {Ma}}, \bibinfo {author}
  {\bibfnamefont {J.-P.}\ \bibnamefont {Hu}}, \bibinfo {author} {\bibfnamefont
  {P.}~\bibnamefont {Chen}}, \ and\ \bibinfo {author} {\bibfnamefont {Q.-K.}\
  \bibnamefont {Xue}},\ }\href {\doibase 10.1038/nphys2155} {\bibfield
  {journal} {\bibinfo  {journal} {Nature Physics}\ }\textbf {\bibinfo {volume}
  {8}},\ \bibinfo {pages} {126} (\bibinfo {year} {2012})}\BibitemShut {NoStop}%
\bibitem [{\citenamefont {Lazarevi\ifmmode~\acute{c}\else \'{c}\fi{}}\ \emph
  {et~al.}(2012)\citenamefont {Lazarevi\ifmmode~\acute{c}\else \'{c}\fi{}},
  \citenamefont {Abeykoon}, \citenamefont {Stephens}, \citenamefont {Lei},
  \citenamefont {Bozin}, \citenamefont {Petrovic},\ and\ \citenamefont
  {Popovi\ifmmode~\acute{c}\else \'{c}\fi{}}}]{Lazarevic_PRB2012}%
  \BibitemOpen
  \bibfield  {author} {\bibinfo {author} {\bibfnamefont {N.}~\bibnamefont
  {Lazarevi\ifmmode~\acute{c}\else \'{c}\fi{}}}, \bibinfo {author}
  {\bibfnamefont {M.}~\bibnamefont {Abeykoon}}, \bibinfo {author}
  {\bibfnamefont {P.~W.}\ \bibnamefont {Stephens}}, \bibinfo {author}
  {\bibfnamefont {H.}~\bibnamefont {Lei}}, \bibinfo {author} {\bibfnamefont
  {E.~S.}\ \bibnamefont {Bozin}}, \bibinfo {author} {\bibfnamefont
  {C.}~\bibnamefont {Petrovic}}, \ and\ \bibinfo {author} {\bibfnamefont
  {Z.~V.}\ \bibnamefont {Popovi\ifmmode~\acute{c}\else \'{c}\fi{}}},\ }\href
  {\doibase 10.1103/PhysRevB.86.054503} {\bibfield  {journal} {\bibinfo
  {journal} {Phys. Rev. B}\ }\textbf {\bibinfo {volume} {86}},\ \bibinfo
  {pages} {054503} (\bibinfo {year} {2012})}\BibitemShut {NoStop}%
\bibitem [{\citenamefont {Badding}\ \emph {et~al.}(1992)\citenamefont
  {Badding}, \citenamefont {Li}, \citenamefont {DiSalvo}, \citenamefont
  {Zhou},\ and\ \citenamefont {Edwards}}]{Badding_JSCS1992}%
  \BibitemOpen
  \bibfield  {author} {\bibinfo {author} {\bibfnamefont {M.}~\bibnamefont
  {Badding}}, \bibinfo {author} {\bibfnamefont {J.}~\bibnamefont {Li}},
  \bibinfo {author} {\bibfnamefont {F.}~\bibnamefont {DiSalvo}}, \bibinfo
  {author} {\bibfnamefont {W.}~\bibnamefont {Zhou}}, \ and\ \bibinfo {author}
  {\bibfnamefont {P.}~\bibnamefont {Edwards}},\ }\href {\doibase
  http://dx.doi.org/10.1016/0022-4596(92)90106-6} {\bibfield  {journal}
  {\bibinfo  {journal} {Journal of Solid State Chemistry}\ }\textbf {\bibinfo
  {volume} {100}},\ \bibinfo {pages} {313 } (\bibinfo {year}
  {1992})}\BibitemShut {NoStop}%
\bibitem [{\citenamefont {Neuhausen}\ \emph {et~al.}(1993)\citenamefont
  {Neuhausen}, \citenamefont {Potthoff}, \citenamefont {Tremel}, \citenamefont
  {Ensling}, \citenamefont {G\"utlich},\ and\ \citenamefont
  {Kremer}}]{Neuhausen_ZNB1993}%
  \BibitemOpen
  \bibfield  {author} {\bibinfo {author} {\bibfnamefont {J.}~\bibnamefont
  {Neuhausen}}, \bibinfo {author} {\bibfnamefont {E.}~\bibnamefont {Potthoff}},
  \bibinfo {author} {\bibfnamefont {W.}~\bibnamefont {Tremel}}, \bibinfo
  {author} {\bibfnamefont {J.}~\bibnamefont {Ensling}}, \bibinfo {author}
  {\bibfnamefont {P.}~\bibnamefont {G\"utlich}}, \ and\ \bibinfo {author}
  {\bibfnamefont {R.}~\bibnamefont {Kremer}},\ }\href {\doibase
  10.1515/znb-1993-0615} {\bibfield  {journal} {\bibinfo  {journal}
  {Zeitschrift f\"ur Naturforschung B}\ }\textbf {\bibinfo {volume} {48}},\
  \bibinfo {pages} {797} (\bibinfo {year} {1993})}\BibitemShut {NoStop}%
\bibitem [{\citenamefont {Perez~Vicente}\ \emph {et~al.}(1998)\citenamefont
  {Perez~Vicente}, \citenamefont {Womes}, \citenamefont {Jumas}, \citenamefont
  {Sanchez},\ and\ \citenamefont {Tirado}}]{Perez_JPCB1998}%
  \BibitemOpen
  \bibfield  {author} {\bibinfo {author} {\bibfnamefont {C.}~\bibnamefont
  {Perez~Vicente}}, \bibinfo {author} {\bibfnamefont {M.}~\bibnamefont
  {Womes}}, \bibinfo {author} {\bibfnamefont {J.~C.}\ \bibnamefont {Jumas}},
  \bibinfo {author} {\bibfnamefont {L.}~\bibnamefont {Sanchez}}, \ and\
  \bibinfo {author} {\bibfnamefont {J.~L.}\ \bibnamefont {Tirado}},\ }\href
  {\doibase 10.1021/jp982070m} {\bibfield  {journal} {\bibinfo  {journal} {The
  Journal of Physical Chemistry B}\ }\textbf {\bibinfo {volume} {102}},\
  \bibinfo {pages} {8712} (\bibinfo {year} {1998})}\BibitemShut {NoStop}%
\bibitem [{\citenamefont {Liu}\ \emph {et~al.}(2011)\citenamefont {Liu},
  \citenamefont {Zhang}, \citenamefont {Cheng}, \citenamefont {Yan},
  \citenamefont {Xiang}, \citenamefont {Ying}, \citenamefont {Wang},
  \citenamefont {Wang}, \citenamefont {Ye}, \citenamefont {Luo},\ and\
  \citenamefont {Chen}}]{Liu_PRB2011}%
  \BibitemOpen
  \bibfield  {author} {\bibinfo {author} {\bibfnamefont {R.~H.}\ \bibnamefont
  {Liu}}, \bibinfo {author} {\bibfnamefont {M.}~\bibnamefont {Zhang}}, \bibinfo
  {author} {\bibfnamefont {P.}~\bibnamefont {Cheng}}, \bibinfo {author}
  {\bibfnamefont {Y.~J.}\ \bibnamefont {Yan}}, \bibinfo {author} {\bibfnamefont
  {Z.~J.}\ \bibnamefont {Xiang}}, \bibinfo {author} {\bibfnamefont {J.~J.}\
  \bibnamefont {Ying}}, \bibinfo {author} {\bibfnamefont {X.~F.}\ \bibnamefont
  {Wang}}, \bibinfo {author} {\bibfnamefont {A.~F.}\ \bibnamefont {Wang}},
  \bibinfo {author} {\bibfnamefont {G.~J.}\ \bibnamefont {Ye}}, \bibinfo
  {author} {\bibfnamefont {X.~G.}\ \bibnamefont {Luo}}, \ and\ \bibinfo
  {author} {\bibfnamefont {X.~H.}\ \bibnamefont {Chen}},\ }\href {\doibase
  10.1103/PhysRevB.84.184432} {\bibfield  {journal} {\bibinfo  {journal} {Phys.
  Rev. B}\ }\textbf {\bibinfo {volume} {84}},\ \bibinfo {pages} {184432}
  (\bibinfo {year} {2011})}\BibitemShut {NoStop}%
\bibitem [{\citenamefont {Ke}\ \emph {et~al.}(2012)\citenamefont {Ke},
  \citenamefont {Qian}, \citenamefont {Cao}, \citenamefont {Hu}, \citenamefont
  {Wang},\ and\ \citenamefont {Mao}}]{Ke_PRB2012}%
  \BibitemOpen
  \bibfield  {author} {\bibinfo {author} {\bibfnamefont {X.}~\bibnamefont
  {Ke}}, \bibinfo {author} {\bibfnamefont {B.}~\bibnamefont {Qian}}, \bibinfo
  {author} {\bibfnamefont {H.}~\bibnamefont {Cao}}, \bibinfo {author}
  {\bibfnamefont {J.}~\bibnamefont {Hu}}, \bibinfo {author} {\bibfnamefont
  {G.~C.}\ \bibnamefont {Wang}}, \ and\ \bibinfo {author} {\bibfnamefont
  {Z.~Q.}\ \bibnamefont {Mao}},\ }\href {\doibase 10.1103/PhysRevB.85.214404}
  {\bibfield  {journal} {\bibinfo  {journal} {Phys. Rev. B}\ }\textbf {\bibinfo
  {volume} {85}},\ \bibinfo {pages} {214404} (\bibinfo {year}
  {2012})}\BibitemShut {NoStop}%
\bibitem [{\citenamefont {Dagotto}(2013)}]{Dagotto_RMP2013}%
  \BibitemOpen
  \bibfield  {author} {\bibinfo {author} {\bibfnamefont {E.}~\bibnamefont
  {Dagotto}},\ }\href {\doibase 10.1103/RevModPhys.85.849} {\bibfield
  {journal} {\bibinfo  {journal} {Rev. Mod. Phys.}\ }\textbf {\bibinfo {volume}
  {85}},\ \bibinfo {pages} {849} (\bibinfo {year} {2013})}\BibitemShut
  {NoStop}%
\bibitem [{\citenamefont {Li}\ \emph {et~al.}(2014)\citenamefont {Li},
  \citenamefont {Setty}, \citenamefont {Chen},\ and\ \citenamefont
  {Hu}}]{Li_FP2014}%
  \BibitemOpen
  \bibfield  {author} {\bibinfo {author} {\bibfnamefont {W.}~\bibnamefont
  {Li}}, \bibinfo {author} {\bibfnamefont {C.}~\bibnamefont {Setty}}, \bibinfo
  {author} {\bibfnamefont {X.~H.}\ \bibnamefont {Chen}}, \ and\ \bibinfo
  {author} {\bibfnamefont {J.}~\bibnamefont {Hu}},\ }\href {\doibase
  10.1007/s11467-014-0428-y} {\bibfield  {journal} {\bibinfo  {journal}
  {Frontiers of Physics}\ }\textbf {\bibinfo {volume} {9}},\ \bibinfo {pages}
  {465} (\bibinfo {year} {2014})}\BibitemShut {NoStop}%
\bibitem [{\citenamefont {Dagotto}\ \emph {et~al.}(1992)\citenamefont
  {Dagotto}, \citenamefont {Riera},\ and\ \citenamefont
  {Scalapino}}]{Dagotto_PRB1992}%
  \BibitemOpen
  \bibfield  {author} {\bibinfo {author} {\bibfnamefont {E.}~\bibnamefont
  {Dagotto}}, \bibinfo {author} {\bibfnamefont {J.}~\bibnamefont {Riera}}, \
  and\ \bibinfo {author} {\bibfnamefont {D.}~\bibnamefont {Scalapino}},\ }\href
  {\doibase 10.1103/PhysRevB.45.5744} {\bibfield  {journal} {\bibinfo
  {journal} {Phys. Rev. B}\ }\textbf {\bibinfo {volume} {45}},\ \bibinfo
  {pages} {5744} (\bibinfo {year} {1992})}\BibitemShut {NoStop}%
\bibitem [{\citenamefont {Dagotto}\ and\ \citenamefont
  {Rice}(1996)}]{Dagotto_SCI1996}%
  \BibitemOpen
  \bibfield  {author} {\bibinfo {author} {\bibfnamefont {E.}~\bibnamefont
  {Dagotto}}\ and\ \bibinfo {author} {\bibfnamefont {T.~M.}\ \bibnamefont
  {Rice}},\ }\href {\doibase 10.1126/science.271.5249.618} {\bibfield
  {journal} {\bibinfo  {journal} {Science}\ }\textbf {\bibinfo {volume}
  {271}},\ \bibinfo {pages} {618} (\bibinfo {year} {1996})}\BibitemShut
  {NoStop}%
\bibitem [{\citenamefont {Dagotto}(1999)}]{Dagotto_RPP1999}%
  \BibitemOpen
  \bibfield  {author} {\bibinfo {author} {\bibfnamefont {E.}~\bibnamefont
  {Dagotto}},\ }\href {http://stacks.iop.org/0034-4885/62/i=11/a=202}
  {\bibfield  {journal} {\bibinfo  {journal} {Reports on Progress in Physics}\
  }\textbf {\bibinfo {volume} {62}},\ \bibinfo {pages} {1525} (\bibinfo {year}
  {1999})}\BibitemShut {NoStop}%
\bibitem [{\citenamefont {Uehara}\ \emph {et~al.}(1996)\citenamefont {Uehara},
  \citenamefont {Nagata}, \citenamefont {Akimitsu}, \citenamefont {Takahashi},
  \citenamefont {M\^{o}ri},\ and\ \citenamefont {Kinoshita}}]{Uehara_JPSJ1996}%
  \BibitemOpen
  \bibfield  {author} {\bibinfo {author} {\bibfnamefont {M.}~\bibnamefont
  {Uehara}}, \bibinfo {author} {\bibfnamefont {T.}~\bibnamefont {Nagata}},
  \bibinfo {author} {\bibfnamefont {J.}~\bibnamefont {Akimitsu}}, \bibinfo
  {author} {\bibfnamefont {H.}~\bibnamefont {Takahashi}}, \bibinfo {author}
  {\bibfnamefont {N.}~\bibnamefont {M\^{o}ri}}, \ and\ \bibinfo {author}
  {\bibfnamefont {K.}~\bibnamefont {Kinoshita}},\ }\href {\doibase
  10.1143/JPSJ.65.2764} {\bibfield  {journal} {\bibinfo  {journal} {Journal of
  the Physical Society of Japan}\ }\textbf {\bibinfo {volume} {65}},\ \bibinfo
  {pages} {2764} (\bibinfo {year} {1996})}\BibitemShut {NoStop}%
\bibitem [{\citenamefont {Hisada}\ \emph {et~al.}(2014)\citenamefont {Hisada},
  \citenamefont {Matsubayashi}, \citenamefont {Uwatoko}, \citenamefont
  {Fujiwara}, \citenamefont {Deng}, \citenamefont {Pomjakushina}, \citenamefont
  {Conder}, \citenamefont {Radheep}, \citenamefont {Thiyagarajan},
  \citenamefont {Esakkimuthu},\ and\ \citenamefont
  {Arumugam}}]{Hisada_JPSJ2014}%
  \BibitemOpen
  \bibfield  {author} {\bibinfo {author} {\bibfnamefont {A.}~\bibnamefont
  {Hisada}}, \bibinfo {author} {\bibfnamefont {K.}~\bibnamefont
  {Matsubayashi}}, \bibinfo {author} {\bibfnamefont {Y.}~\bibnamefont
  {Uwatoko}}, \bibinfo {author} {\bibfnamefont {N.}~\bibnamefont {Fujiwara}},
  \bibinfo {author} {\bibfnamefont {G.}~\bibnamefont {Deng}}, \bibinfo {author}
  {\bibfnamefont {E.}~\bibnamefont {Pomjakushina}}, \bibinfo {author}
  {\bibfnamefont {K.}~\bibnamefont {Conder}}, \bibinfo {author} {\bibfnamefont
  {D.~M.}\ \bibnamefont {Radheep}}, \bibinfo {author} {\bibfnamefont
  {R.}~\bibnamefont {Thiyagarajan}}, \bibinfo {author} {\bibfnamefont
  {S.}~\bibnamefont {Esakkimuthu}}, \ and\ \bibinfo {author} {\bibfnamefont
  {S.}~\bibnamefont {Arumugam}},\ }\href {\doibase 10.7566/JPSJ.83.073703}
  {\bibfield  {journal} {\bibinfo  {journal} {Journal of the Physical Society
  of Japan}\ }\textbf {\bibinfo {volume} {83}},\ \bibinfo {pages} {073703}
  (\bibinfo {year} {2014})}\BibitemShut {NoStop}%
\bibitem [{\citenamefont {Takahashi}\ \emph {et~al.}(2014)\citenamefont
  {Takahashi}, \citenamefont {Sugimoto}, \citenamefont {Nambu}, \citenamefont
  {Yamauchi}, \citenamefont {Hirata}, \citenamefont {Kawakami}, \citenamefont
  {Avdeev}, \citenamefont {Matsubayashi}, \citenamefont {Du}, \citenamefont
  {Kawashima}, \citenamefont {Soeda}, \citenamefont {Nakano}, \citenamefont
  {Uwatoko}, \citenamefont {Ueda}, \citenamefont {Sato},\ and\ \citenamefont
  {Ohgushi}}]{Takahashi_NMat2014}%
  \BibitemOpen
  \bibfield  {author} {\bibinfo {author} {\bibfnamefont {H.}~\bibnamefont
  {Takahashi}}, \bibinfo {author} {\bibfnamefont {A.}~\bibnamefont {Sugimoto}},
  \bibinfo {author} {\bibfnamefont {Y.}~\bibnamefont {Nambu}}, \bibinfo
  {author} {\bibfnamefont {T.}~\bibnamefont {Yamauchi}}, \bibinfo {author}
  {\bibfnamefont {Y.}~\bibnamefont {Hirata}}, \bibinfo {author} {\bibfnamefont
  {T.}~\bibnamefont {Kawakami}}, \bibinfo {author} {\bibfnamefont
  {M.}~\bibnamefont {Avdeev}}, \bibinfo {author} {\bibfnamefont
  {K.}~\bibnamefont {Matsubayashi}}, \bibinfo {author} {\bibfnamefont
  {F.}~\bibnamefont {Du}}, \bibinfo {author} {\bibfnamefont {C.}~\bibnamefont
  {Kawashima}}, \bibinfo {author} {\bibfnamefont {H.}~\bibnamefont {Soeda}},
  \bibinfo {author} {\bibfnamefont {S.}~\bibnamefont {Nakano}}, \bibinfo
  {author} {\bibfnamefont {Y.}~\bibnamefont {Uwatoko}}, \bibinfo {author}
  {\bibfnamefont {Y.}~\bibnamefont {Ueda}}, \bibinfo {author} {\bibfnamefont
  {T.~J.}\ \bibnamefont {Sato}}, \ and\ \bibinfo {author} {\bibfnamefont
  {K.}~\bibnamefont {Ohgushi}},\ }\href {\doibase 10.1038/nmat4351} {\bibfield
  {journal} {\bibinfo  {journal} {Nat. Mater.}\ }\textbf {\bibinfo {volume}
  {14}},\ \bibinfo {pages} {1008} (\bibinfo {year} {2014})}\BibitemShut
  {NoStop}%
\bibitem [{\citenamefont {Zhang}\ and\ \citenamefont
  {Zhang}(2013)}]{Zhang_CPB2013}%
  \BibitemOpen
  \bibfield  {author} {\bibinfo {author} {\bibfnamefont {A.-M.}\ \bibnamefont
  {Zhang}}\ and\ \bibinfo {author} {\bibfnamefont {Q.-M.}\ \bibnamefont
  {Zhang}},\ }\href {http://stacks.iop.org/1674-1056/22/i=8/a=087103}
  {\bibfield  {journal} {\bibinfo  {journal} {Chin. Phys. B}\ }\textbf
  {\bibinfo {volume} {22}},\ \bibinfo {pages} {087103} (\bibinfo {year}
  {2013})}\BibitemShut {NoStop}%
\bibitem [{\citenamefont {Opa{\v{c}}i{\'c}}\ and\ \citenamefont
  {Lazarevi{\'c}}(2017)}]{opavcic:2017lattice}%
  \BibitemOpen
  \bibfield  {author} {\bibinfo {author} {\bibfnamefont {M.}~\bibnamefont
  {Opa{\v{c}}i{\'c}}}\ and\ \bibinfo {author} {\bibfnamefont {N.}~\bibnamefont
  {Lazarevi{\'c}}},\ }\href@noop {} {\bibfield  {journal} {\bibinfo  {journal}
  {J. Serb. Chem. Soc.}\ } (\bibinfo {year} {2017})}\BibitemShut {NoStop}%
\bibitem [{\citenamefont {Zhang}\ \emph {et~al.}(2009)\citenamefont {Zhang},
  \citenamefont {Fujita}, \citenamefont {Chen}, \citenamefont {Feng},
  \citenamefont {Maekawa},\ and\ \citenamefont {Chen}}]{Zhang_PRB2009}%
  \BibitemOpen
  \bibfield  {author} {\bibinfo {author} {\bibfnamefont {L.}~\bibnamefont
  {Zhang}}, \bibinfo {author} {\bibfnamefont {T.}~\bibnamefont {Fujita}},
  \bibinfo {author} {\bibfnamefont {F.}~\bibnamefont {Chen}}, \bibinfo {author}
  {\bibfnamefont {D.~L.}\ \bibnamefont {Feng}}, \bibinfo {author}
  {\bibfnamefont {S.}~\bibnamefont {Maekawa}}, \ and\ \bibinfo {author}
  {\bibfnamefont {M.~W.}\ \bibnamefont {Chen}},\ }\href {\doibase
  10.1103/PhysRevB.79.052507} {\bibfield  {journal} {\bibinfo  {journal} {Phys.
  Rev. B}\ }\textbf {\bibinfo {volume} {79}},\ \bibinfo {pages} {052507}
  (\bibinfo {year} {2009})}\BibitemShut {NoStop}%
\bibitem [{\citenamefont {Gallais}\ \emph {et~al.}(2008)\citenamefont
  {Gallais}, \citenamefont {Sacuto}, \citenamefont {Cazayous}, \citenamefont
  {Cheng}, \citenamefont {Fang},\ and\ \citenamefont {Wen}}]{Gallais_PRB2008}%
  \BibitemOpen
  \bibfield  {author} {\bibinfo {author} {\bibfnamefont {Y.}~\bibnamefont
  {Gallais}}, \bibinfo {author} {\bibfnamefont {A.}~\bibnamefont {Sacuto}},
  \bibinfo {author} {\bibfnamefont {M.}~\bibnamefont {Cazayous}}, \bibinfo
  {author} {\bibfnamefont {P.}~\bibnamefont {Cheng}}, \bibinfo {author}
  {\bibfnamefont {L.}~\bibnamefont {Fang}}, \ and\ \bibinfo {author}
  {\bibfnamefont {H.~H.}\ \bibnamefont {Wen}},\ }\href {\doibase
  10.1103/PhysRevB.78.132509} {\bibfield  {journal} {\bibinfo  {journal} {Phys.
  Rev. B}\ }\textbf {\bibinfo {volume} {78}},\ \bibinfo {pages} {132509}
  (\bibinfo {year} {2008})}\BibitemShut {NoStop}%
\bibitem [{\citenamefont {Gnezdilov}\ \emph {et~al.}(2013)\citenamefont
  {Gnezdilov}, \citenamefont {Pashkevich}, \citenamefont {Lemmens},
  \citenamefont {Wulferding}, \citenamefont {Shevtsova}, \citenamefont {Gusev},
  \citenamefont {Chareev},\ and\ \citenamefont {Vasiliev}}]{Gnezdilov_PRB2013}%
  \BibitemOpen
  \bibfield  {author} {\bibinfo {author} {\bibfnamefont {V.}~\bibnamefont
  {Gnezdilov}}, \bibinfo {author} {\bibfnamefont {Y.~G.}\ \bibnamefont
  {Pashkevich}}, \bibinfo {author} {\bibfnamefont {P.}~\bibnamefont {Lemmens}},
  \bibinfo {author} {\bibfnamefont {D.}~\bibnamefont {Wulferding}}, \bibinfo
  {author} {\bibfnamefont {T.}~\bibnamefont {Shevtsova}}, \bibinfo {author}
  {\bibfnamefont {A.}~\bibnamefont {Gusev}}, \bibinfo {author} {\bibfnamefont
  {D.}~\bibnamefont {Chareev}}, \ and\ \bibinfo {author} {\bibfnamefont
  {A.}~\bibnamefont {Vasiliev}},\ }\href {\doibase 10.1103/PhysRevB.87.144508}
  {\bibfield  {journal} {\bibinfo  {journal} {Phys. Rev. B}\ }\textbf {\bibinfo
  {volume} {87}},\ \bibinfo {pages} {144508} (\bibinfo {year}
  {2013})}\BibitemShut {NoStop}%
\bibitem [{\citenamefont {Litvinchuk}\ \emph {et~al.}(2011)\citenamefont
  {Litvinchuk}, \citenamefont {Lv},\ and\ \citenamefont
  {Chu}}]{Litvinchuk_PRB2011}%
  \BibitemOpen
  \bibfield  {author} {\bibinfo {author} {\bibfnamefont {A.~P.}\ \bibnamefont
  {Litvinchuk}}, \bibinfo {author} {\bibfnamefont {B.}~\bibnamefont {Lv}}, \
  and\ \bibinfo {author} {\bibfnamefont {C.~W.}\ \bibnamefont {Chu}},\ }\href
  {\doibase 10.1103/PhysRevB.84.092504} {\bibfield  {journal} {\bibinfo
  {journal} {Phys. Rev. B}\ }\textbf {\bibinfo {volume} {84}},\ \bibinfo
  {pages} {092504} (\bibinfo {year} {2011})}\BibitemShut {NoStop}%
\bibitem [{\citenamefont {Choi}\ \emph {et~al.}(2010)\citenamefont {Choi},
  \citenamefont {Lemmens}, \citenamefont {Eremin}, \citenamefont {Zwicknagl},
  \citenamefont {Berger}, \citenamefont {Sun}, \citenamefont {Sun},\ and\
  \citenamefont {Lin}}]{Choi_JPCM2010}%
  \BibitemOpen
  \bibfield  {author} {\bibinfo {author} {\bibfnamefont {K.-Y.}\ \bibnamefont
  {Choi}}, \bibinfo {author} {\bibfnamefont {P.}~\bibnamefont {Lemmens}},
  \bibinfo {author} {\bibfnamefont {I.}~\bibnamefont {Eremin}}, \bibinfo
  {author} {\bibfnamefont {G.}~\bibnamefont {Zwicknagl}}, \bibinfo {author}
  {\bibfnamefont {H.}~\bibnamefont {Berger}}, \bibinfo {author} {\bibfnamefont
  {G.~L.}\ \bibnamefont {Sun}}, \bibinfo {author} {\bibfnamefont {D.~L.}\
  \bibnamefont {Sun}}, \ and\ \bibinfo {author} {\bibfnamefont {C.~T.}\
  \bibnamefont {Lin}},\ }\href
  {http://stacks.iop.org/0953-8984/22/i=11/a=115802} {\bibfield  {journal}
  {\bibinfo  {journal} {Journal of Physics: Condensed Matter}\ }\textbf
  {\bibinfo {volume} {22}},\ \bibinfo {pages} {115802} (\bibinfo {year}
  {2010})}\BibitemShut {NoStop}%
\bibitem [{\citenamefont {Baroni}\ \emph {et~al.}(2001)\citenamefont {Baroni},
  \citenamefont {de~Gironcoli}, \citenamefont {Dal~Corso},\ and\ \citenamefont
  {Giannozzi}}]{Baroni_RevModPhys2001}%
  \BibitemOpen
  \bibfield  {author} {\bibinfo {author} {\bibfnamefont {S.}~\bibnamefont
  {Baroni}}, \bibinfo {author} {\bibfnamefont {S.}~\bibnamefont
  {de~Gironcoli}}, \bibinfo {author} {\bibfnamefont {A.}~\bibnamefont
  {Dal~Corso}}, \ and\ \bibinfo {author} {\bibfnamefont {P.}~\bibnamefont
  {Giannozzi}},\ }\href {\doibase 10.1103/RevModPhys.73.515} {\bibfield
  {journal} {\bibinfo  {journal} {Rev. Mod. Phys.}\ }\textbf {\bibinfo {volume}
  {73}},\ \bibinfo {pages} {515} (\bibinfo {year} {2001})}\BibitemShut
  {NoStop}%
\bibitem [{\citenamefont {Gianozzi}\ \emph {et~al.}(2009)\citenamefont
  {Gianozzi}, \citenamefont {Baroni}, \citenamefont {Bonini}, \citenamefont
  {Calandra}, \citenamefont {Car}, \citenamefont {Cavazzoni}, \citenamefont
  {Ceresoli}, \citenamefont {Chiarotti}, \citenamefont {Cococcioni},
  \citenamefont {Dabo}, \citenamefont {Corso}, \citenamefont {de~Gironcoli},
  \citenamefont {Fabris}, \citenamefont {Fratesi}, \citenamefont {Gebauer},
  \citenamefont {Gerstmann}, \citenamefont {Gougoussis}, \citenamefont
  {Kokalj}, \citenamefont {Lazzeri}, \citenamefont {Martin-Samos},
  \citenamefont {Marzari}, \citenamefont {Mauri}, \citenamefont {Mazzarello},
  \citenamefont {Paolini}, \citenamefont {Pasquarello}, \citenamefont
  {Paulatto}, \citenamefont {Sbraccia}, \citenamefont {Scandolo}, \citenamefont
  {Sclauzero}, \citenamefont {Seitsonen}, \citenamefont {Smogunov},
  \citenamefont {Umari},\ and\ \citenamefont
  {Wentzcovitch}}]{Gianozzi_JPCM2009}%
  \BibitemOpen
  \bibfield  {author} {\bibinfo {author} {\bibfnamefont {P.}~\bibnamefont
  {Gianozzi}}, \bibinfo {author} {\bibfnamefont {S.}~\bibnamefont {Baroni}},
  \bibinfo {author} {\bibfnamefont {N.}~\bibnamefont {Bonini}}, \bibinfo
  {author} {\bibfnamefont {M.}~\bibnamefont {Calandra}}, \bibinfo {author}
  {\bibfnamefont {R.}~\bibnamefont {Car}}, \bibinfo {author} {\bibfnamefont
  {C.}~\bibnamefont {Cavazzoni}}, \bibinfo {author} {\bibfnamefont
  {D.}~\bibnamefont {Ceresoli}}, \bibinfo {author} {\bibfnamefont {G.~L.}\
  \bibnamefont {Chiarotti}}, \bibinfo {author} {\bibfnamefont {M.}~\bibnamefont
  {Cococcioni}}, \bibinfo {author} {\bibfnamefont {I.}~\bibnamefont {Dabo}},
  \bibinfo {author} {\bibfnamefont {A.~D.}\ \bibnamefont {Corso}}, \bibinfo
  {author} {\bibfnamefont {S.}~\bibnamefont {de~Gironcoli}}, \bibinfo {author}
  {\bibfnamefont {S.}~\bibnamefont {Fabris}}, \bibinfo {author} {\bibfnamefont
  {G.}~\bibnamefont {Fratesi}}, \bibinfo {author} {\bibfnamefont
  {R.}~\bibnamefont {Gebauer}}, \bibinfo {author} {\bibfnamefont
  {U.}~\bibnamefont {Gerstmann}}, \bibinfo {author} {\bibfnamefont
  {C.}~\bibnamefont {Gougoussis}}, \bibinfo {author} {\bibfnamefont
  {A.}~\bibnamefont {Kokalj}}, \bibinfo {author} {\bibfnamefont
  {M.}~\bibnamefont {Lazzeri}}, \bibinfo {author} {\bibfnamefont
  {L.}~\bibnamefont {Martin-Samos}}, \bibinfo {author} {\bibfnamefont
  {N.}~\bibnamefont {Marzari}}, \bibinfo {author} {\bibfnamefont
  {F.}~\bibnamefont {Mauri}}, \bibinfo {author} {\bibfnamefont
  {R.}~\bibnamefont {Mazzarello}}, \bibinfo {author} {\bibfnamefont
  {S.}~\bibnamefont {Paolini}}, \bibinfo {author} {\bibfnamefont
  {A.}~\bibnamefont {Pasquarello}}, \bibinfo {author} {\bibfnamefont
  {L.}~\bibnamefont {Paulatto}}, \bibinfo {author} {\bibfnamefont
  {C.}~\bibnamefont {Sbraccia}}, \bibinfo {author} {\bibfnamefont
  {S.}~\bibnamefont {Scandolo}}, \bibinfo {author} {\bibfnamefont
  {G.}~\bibnamefont {Sclauzero}}, \bibinfo {author} {\bibfnamefont {A.~P.}\
  \bibnamefont {Seitsonen}}, \bibinfo {author} {\bibfnamefont {A.}~\bibnamefont
  {Smogunov}}, \bibinfo {author} {\bibfnamefont {P.}~\bibnamefont {Umari}}, \
  and\ \bibinfo {author} {\bibfnamefont {R.~M.}\ \bibnamefont {Wentzcovitch}},\
  }\href {http://stacks.iop.org/0953-8984/21/i=39/a=395502} {\bibfield
  {journal} {\bibinfo  {journal} {Journal of Physics: Condensed Matter}\
  }\textbf {\bibinfo {volume} {21}},\ \bibinfo {pages} {395502} (\bibinfo
  {year} {2009})}\BibitemShut {NoStop}%
\bibitem [{\citenamefont {Rousseau}\ \emph {et~al.}(1981)\citenamefont
  {Rousseau}, \citenamefont {Bauman},\ and\ \citenamefont
  {Porto}}]{Porto:1981}%
  \BibitemOpen
  \bibfield  {author} {\bibinfo {author} {\bibfnamefont {D.~L.}\ \bibnamefont
  {Rousseau}}, \bibinfo {author} {\bibfnamefont {R.~P.}\ \bibnamefont
  {Bauman}}, \ and\ \bibinfo {author} {\bibfnamefont {S.~P.~S.}\ \bibnamefont
  {Porto}},\ }\href {\doibase 10.1002/jrs.1250100152} {\bibfield  {journal}
  {\bibinfo  {journal} {Journal of Raman Spectroscopy}\ }\textbf {\bibinfo
  {volume} {10}},\ \bibinfo {pages} {253} (\bibinfo {year} {1981})}\BibitemShut
  {NoStop}%
\bibitem [{\citenamefont {Min}\ \emph {et~al.}(2015)\citenamefont {Min},
  \citenamefont {Li-Min}, \citenamefont {Rui}, \citenamefont {Qing-Qin},
  \citenamefont {Fei}, \citenamefont {Zi-Rong}, \citenamefont {Yan},
  \citenamefont {Su-Di}, \citenamefont {Miao}, \citenamefont {Rong-Hua},
  \citenamefont {Arita}, \citenamefont {Shimada}, \citenamefont {Namatame},
  \citenamefont {Taniguchi}, \citenamefont {Matsunami}, \citenamefont {Kimura},
  \citenamefont {Ming}, \citenamefont {Xian-Hui}, \citenamefont {Wei-Guo},
  \citenamefont {Wei}, \citenamefont {Bin-Ping},\ and\ \citenamefont
  {Dong-Lai}}]{Min_CPL2015}%
  \BibitemOpen
  \bibfield  {author} {\bibinfo {author} {\bibfnamefont {X.}~\bibnamefont
  {Min}}, \bibinfo {author} {\bibfnamefont {W.}~\bibnamefont {Li-Min}},
  \bibinfo {author} {\bibfnamefont {P.}~\bibnamefont {Rui}}, \bibinfo {author}
  {\bibfnamefont {G.}~\bibnamefont {Qing-Qin}}, \bibinfo {author}
  {\bibfnamefont {C.}~\bibnamefont {Fei}}, \bibinfo {author} {\bibfnamefont
  {Y.}~\bibnamefont {Zi-Rong}}, \bibinfo {author} {\bibfnamefont
  {Z.}~\bibnamefont {Yan}}, \bibinfo {author} {\bibfnamefont {C.}~\bibnamefont
  {Su-Di}}, \bibinfo {author} {\bibfnamefont {X.}~\bibnamefont {Miao}},
  \bibinfo {author} {\bibfnamefont {L.}~\bibnamefont {Rong-Hua}}, \bibinfo
  {author} {\bibfnamefont {M.}~\bibnamefont {Arita}}, \bibinfo {author}
  {\bibfnamefont {K.}~\bibnamefont {Shimada}}, \bibinfo {author} {\bibfnamefont
  {H.}~\bibnamefont {Namatame}}, \bibinfo {author} {\bibfnamefont
  {M.}~\bibnamefont {Taniguchi}}, \bibinfo {author} {\bibfnamefont
  {M.}~\bibnamefont {Matsunami}}, \bibinfo {author} {\bibfnamefont
  {S.}~\bibnamefont {Kimura}}, \bibinfo {author} {\bibfnamefont
  {S.}~\bibnamefont {Ming}}, \bibinfo {author} {\bibfnamefont {C.}~\bibnamefont
  {Xian-Hui}}, \bibinfo {author} {\bibfnamefont {Y.}~\bibnamefont {Wei-Guo}},
  \bibinfo {author} {\bibfnamefont {K.}~\bibnamefont {Wei}}, \bibinfo {author}
  {\bibfnamefont {X.}~\bibnamefont {Bin-Ping}}, \ and\ \bibinfo {author}
  {\bibfnamefont {F.}~\bibnamefont {Dong-Lai}},\ }\href
  {http://stacks.iop.org/0256-307X/32/i=2/a=027401} {\bibfield  {journal}
  {\bibinfo  {journal} {Chinese Physics Letters}\ }\textbf {\bibinfo {volume}
  {32}},\ \bibinfo {pages} {027401} (\bibinfo {year} {2015})}\BibitemShut
  {NoStop}%
\bibitem [{\citenamefont {Shuker}\ and\ \citenamefont
  {Gammon}(1970)}]{Shuker_PRL1970}%
  \BibitemOpen
  \bibfield  {author} {\bibinfo {author} {\bibfnamefont {R.}~\bibnamefont
  {Shuker}}\ and\ \bibinfo {author} {\bibfnamefont {R.~W.}\ \bibnamefont
  {Gammon}},\ }\href {\doibase 10.1103/PhysRevLett.25.222} {\bibfield
  {journal} {\bibinfo  {journal} {Phys. Rev. Lett.}\ }\textbf {\bibinfo
  {volume} {25}},\ \bibinfo {pages} {222} (\bibinfo {year} {1970})}\BibitemShut
  {NoStop}%
\bibitem [{\citenamefont {Benassi}\ \emph {et~al.}(1991)\citenamefont
  {Benassi}, \citenamefont {Pilla}, \citenamefont {Mazzacurati}, \citenamefont
  {Montagna}, \citenamefont {Ruocco},\ and\ \citenamefont
  {Signorelli}}]{Benassi_PRB1991}%
  \BibitemOpen
  \bibfield  {author} {\bibinfo {author} {\bibfnamefont {P.}~\bibnamefont
  {Benassi}}, \bibinfo {author} {\bibfnamefont {O.}~\bibnamefont {Pilla}},
  \bibinfo {author} {\bibfnamefont {V.}~\bibnamefont {Mazzacurati}}, \bibinfo
  {author} {\bibfnamefont {M.}~\bibnamefont {Montagna}}, \bibinfo {author}
  {\bibfnamefont {G.}~\bibnamefont {Ruocco}}, \ and\ \bibinfo {author}
  {\bibfnamefont {G.}~\bibnamefont {Signorelli}},\ }\href {\doibase
  10.1103/PhysRevB.44.11734} {\bibfield  {journal} {\bibinfo  {journal} {Phys.
  Rev. B}\ }\textbf {\bibinfo {volume} {44}},\ \bibinfo {pages} {11734}
  (\bibinfo {year} {1991})}\BibitemShut {NoStop}%
\bibitem [{\citenamefont {Ryu}\ \emph {et~al.}(2015)\citenamefont {Ryu},
  \citenamefont {Abeykoon}, \citenamefont {Wang}, \citenamefont {Lei},
  \citenamefont {Lazarevic}, \citenamefont {Warren}, \citenamefont {Bozin},
  \citenamefont {Popovic},\ and\ \citenamefont {Petrovic}}]{Ryu2_PRB2015}%
  \BibitemOpen
  \bibfield  {author} {\bibinfo {author} {\bibfnamefont {H.}~\bibnamefont
  {Ryu}}, \bibinfo {author} {\bibfnamefont {M.}~\bibnamefont {Abeykoon}},
  \bibinfo {author} {\bibfnamefont {K.}~\bibnamefont {Wang}}, \bibinfo {author}
  {\bibfnamefont {H.}~\bibnamefont {Lei}}, \bibinfo {author} {\bibfnamefont
  {N.}~\bibnamefont {Lazarevic}}, \bibinfo {author} {\bibfnamefont {J.~B.}\
  \bibnamefont {Warren}}, \bibinfo {author} {\bibfnamefont {E.~S.}\
  \bibnamefont {Bozin}}, \bibinfo {author} {\bibfnamefont {Z.~V.}\ \bibnamefont
  {Popovic}}, \ and\ \bibinfo {author} {\bibfnamefont {C.}~\bibnamefont
  {Petrovic}},\ }\href {\doibase 10.1103/PhysRevB.91.184503} {\bibfield
  {journal} {\bibinfo  {journal} {Phys. Rev. B}\ }\textbf {\bibinfo {volume}
  {91}},\ \bibinfo {pages} {184503} (\bibinfo {year} {2015})}\BibitemShut
  {NoStop}%
\bibitem [{\citenamefont {Zhang}\ \emph {et~al.}(2015)\citenamefont {Zhang},
  \citenamefont {Zhang}, \citenamefont {Dong}, \citenamefont {Lv},
  \citenamefont {Chen}, \citenamefont {Yao}, \citenamefont {Zhang},
  \citenamefont {Gu}, \citenamefont {Zhou}, \citenamefont {Guedes},
  \citenamefont {Yu},\ and\ \citenamefont {Chen}}]{Zhang_AIP2015}%
  \BibitemOpen
  \bibfield  {author} {\bibinfo {author} {\bibfnamefont {B.-B.}\ \bibnamefont
  {Zhang}}, \bibinfo {author} {\bibfnamefont {N.}~\bibnamefont {Zhang}},
  \bibinfo {author} {\bibfnamefont {S.-T.}\ \bibnamefont {Dong}}, \bibinfo
  {author} {\bibfnamefont {Y.}~\bibnamefont {Lv}}, \bibinfo {author}
  {\bibfnamefont {Y.~B.}\ \bibnamefont {Chen}}, \bibinfo {author}
  {\bibfnamefont {S.}~\bibnamefont {Yao}}, \bibinfo {author} {\bibfnamefont
  {S.-T.}\ \bibnamefont {Zhang}}, \bibinfo {author} {\bibfnamefont {Z.-B.}\
  \bibnamefont {Gu}}, \bibinfo {author} {\bibfnamefont {J.}~\bibnamefont
  {Zhou}}, \bibinfo {author} {\bibfnamefont {I.}~\bibnamefont {Guedes}},
  \bibinfo {author} {\bibfnamefont {D.}~\bibnamefont {Yu}}, \ and\ \bibinfo
  {author} {\bibfnamefont {Y.-F.}\ \bibnamefont {Chen}},\ }\href {\doibase
  10.1063/1.4928384} {\bibfield  {journal} {\bibinfo  {journal} {AIP Advances}\
  }\textbf {\bibinfo {volume} {5}},\ \bibinfo {pages} {087111} (\bibinfo {year}
  {2015})}\BibitemShut {NoStop}%
\bibitem [{\citenamefont {Men\'endez}\ and\ \citenamefont
  {Cardona}(1984)}]{Menendez_PRB1984}%
  \BibitemOpen
  \bibfield  {author} {\bibinfo {author} {\bibfnamefont {J.}~\bibnamefont
  {Men\'endez}}\ and\ \bibinfo {author} {\bibfnamefont {M.}~\bibnamefont
  {Cardona}},\ }\href {\doibase 10.1103/PhysRevB.29.2051} {\bibfield  {journal}
  {\bibinfo  {journal} {Phys. Rev. B}\ }\textbf {\bibinfo {volume} {29}},\
  \bibinfo {pages} {2051} (\bibinfo {year} {1984})}\BibitemShut {NoStop}%
\bibitem [{\citenamefont {Eiter}\ \emph {et~al.}(2014)\citenamefont {Eiter},
  \citenamefont {Jaschke}, \citenamefont {Hackl}, \citenamefont {Bauer},
  \citenamefont {Gangl},\ and\ \citenamefont {Pfleiderer}}]{Eiter_PRB2014}%
  \BibitemOpen
  \bibfield  {author} {\bibinfo {author} {\bibfnamefont {H.-M.}\ \bibnamefont
  {Eiter}}, \bibinfo {author} {\bibfnamefont {P.}~\bibnamefont {Jaschke}},
  \bibinfo {author} {\bibfnamefont {R.}~\bibnamefont {Hackl}}, \bibinfo
  {author} {\bibfnamefont {A.}~\bibnamefont {Bauer}}, \bibinfo {author}
  {\bibfnamefont {M.}~\bibnamefont {Gangl}}, \ and\ \bibinfo {author}
  {\bibfnamefont {C.}~\bibnamefont {Pfleiderer}},\ }\href {\doibase
  10.1103/PhysRevB.90.024411} {\bibfield  {journal} {\bibinfo  {journal} {Phys.
  Rev. B}\ }\textbf {\bibinfo {volume} {90}},\ \bibinfo {pages} {024411}
  (\bibinfo {year} {2014})}\BibitemShut {NoStop}%
\bibitem [{\citenamefont {Lazarevi\ifmmode~\acute{c}\else \'{c}\fi{}}\ \emph
  {et~al.}(2013)\citenamefont {Lazarevi\ifmmode~\acute{c}\else \'{c}\fi{}},
  \citenamefont {Radonji\ifmmode~\acute{c}\else \'{c}\fi{}}, \citenamefont
  {\ifmmode \check{S}\else \v{S}\fi{}\ifmmode \acute{c}\else
  \'{c}\fi{}epanovi\ifmmode~\acute{c}\else \'{c}\fi{}}, \citenamefont {Lei},
  \citenamefont {Tanaskovi\ifmmode~\acute{c}\else \'{c}\fi{}}, \citenamefont
  {Petrovic},\ and\ \citenamefont {Popovi\ifmmode~\acute{c}\else
  \'{c}\fi{}}}]{Lazarevic_PRB2013}%
  \BibitemOpen
  \bibfield  {author} {\bibinfo {author} {\bibfnamefont {N.}~\bibnamefont
  {Lazarevi\ifmmode~\acute{c}\else \'{c}\fi{}}}, \bibinfo {author}
  {\bibfnamefont {M.}~\bibnamefont {Radonji\ifmmode~\acute{c}\else
  \'{c}\fi{}}}, \bibinfo {author} {\bibfnamefont {M.}~\bibnamefont {\ifmmode
  \check{S}\else \v{S}\fi{}\ifmmode \acute{c}\else
  \'{c}\fi{}epanovi\ifmmode~\acute{c}\else \'{c}\fi{}}}, \bibinfo {author}
  {\bibfnamefont {H.}~\bibnamefont {Lei}}, \bibinfo {author} {\bibfnamefont
  {D.}~\bibnamefont {Tanaskovi\ifmmode~\acute{c}\else \'{c}\fi{}}}, \bibinfo
  {author} {\bibfnamefont {C.}~\bibnamefont {Petrovic}}, \ and\ \bibinfo
  {author} {\bibfnamefont {Z.~V.}\ \bibnamefont {Popovi\ifmmode~\acute{c}\else
  \'{c}\fi{}}},\ }\href {\doibase 10.1103/PhysRevB.87.144305} {\bibfield
  {journal} {\bibinfo  {journal} {Phys. Rev. B}\ }\textbf {\bibinfo {volume}
  {87}},\ \bibinfo {pages} {144305} (\bibinfo {year} {2013})}\BibitemShut
  {NoStop}%
\bibitem [{\citenamefont {Opa\v{c}i\'c}\ \emph {et~al.}(2015)\citenamefont
  {Opa\v{c}i\'c}, \citenamefont {Lazarevi\'c}, \citenamefont
  {\v{S}\'cepanovi\'c}, \citenamefont {Ryu}, \citenamefont {Lei}, \citenamefont
  {Petrovic},\ and\ \citenamefont {Popovi\'c}}]{Opacic_JPCM2015}%
  \BibitemOpen
  \bibfield  {author} {\bibinfo {author} {\bibfnamefont {M.}~\bibnamefont
  {Opa\v{c}i\'c}}, \bibinfo {author} {\bibfnamefont {N.}~\bibnamefont
  {Lazarevi\'c}}, \bibinfo {author} {\bibfnamefont {M.}~\bibnamefont
  {\v{S}\'cepanovi\'c}}, \bibinfo {author} {\bibfnamefont {H.}~\bibnamefont
  {Ryu}}, \bibinfo {author} {\bibfnamefont {H.}~\bibnamefont {Lei}}, \bibinfo
  {author} {\bibfnamefont {C.}~\bibnamefont {Petrovic}}, \ and\ \bibinfo
  {author} {\bibfnamefont {Z.~V.}\ \bibnamefont {Popovi\'c}},\ }\href
  {http://stacks.iop.org/0953-8984/27/i=48/a=485701} {\bibfield  {journal}
  {\bibinfo  {journal} {Journal of Physics: Condensed Matter}\ }\textbf
  {\bibinfo {volume} {27}},\ \bibinfo {pages} {485701} (\bibinfo {year}
  {2015})}\BibitemShut {NoStop}%
\bibitem [{\citenamefont {Iliev}\ \emph
  {et~al.}(1999{\natexlab{a}})\citenamefont {Iliev}, \citenamefont
  {Litvinchuk}, \citenamefont {Lee}, \citenamefont {Chu}, \citenamefont
  {Barry},\ and\ \citenamefont {Coey}}]{Iliev_PRB1999}%
  \BibitemOpen
  \bibfield  {author} {\bibinfo {author} {\bibfnamefont {M.~N.}\ \bibnamefont
  {Iliev}}, \bibinfo {author} {\bibfnamefont {A.~P.}\ \bibnamefont
  {Litvinchuk}}, \bibinfo {author} {\bibfnamefont {H.-G.}\ \bibnamefont {Lee}},
  \bibinfo {author} {\bibfnamefont {C.~W.}\ \bibnamefont {Chu}}, \bibinfo
  {author} {\bibfnamefont {A.}~\bibnamefont {Barry}}, \ and\ \bibinfo {author}
  {\bibfnamefont {J.~M.~D.}\ \bibnamefont {Coey}},\ }\href {\doibase
  10.1103/PhysRevB.60.33} {\bibfield  {journal} {\bibinfo  {journal} {Phys.
  Rev. B}\ }\textbf {\bibinfo {volume} {60}},\ \bibinfo {pages} {33} (\bibinfo
  {year} {1999}{\natexlab{a}})}\BibitemShut {NoStop}%
\bibitem [{\citenamefont {Lazarevi\ifmmode~\acute{c}\else \'{c}\fi{}}\ \emph
  {et~al.}(2010)\citenamefont {Lazarevi\ifmmode~\acute{c}\else \'{c}\fi{}},
  \citenamefont {Popovi\ifmmode~\acute{c}\else \'{c}\fi{}}, \citenamefont
  {Hu},\ and\ \citenamefont {Petrovic}}]{Lazarevic:2010}%
  \BibitemOpen
  \bibfield  {author} {\bibinfo {author} {\bibfnamefont {N.}~\bibnamefont
  {Lazarevi\ifmmode~\acute{c}\else \'{c}\fi{}}}, \bibinfo {author}
  {\bibfnamefont {Z.~V.}\ \bibnamefont {Popovi\ifmmode~\acute{c}\else
  \'{c}\fi{}}}, \bibinfo {author} {\bibfnamefont {R.}~\bibnamefont {Hu}}, \
  and\ \bibinfo {author} {\bibfnamefont {C.}~\bibnamefont {Petrovic}},\ }\href
  {\doibase 10.1103/PhysRevB.81.144302} {\bibfield  {journal} {\bibinfo
  {journal} {Phys. Rev. B}\ }\textbf {\bibinfo {volume} {81}},\ \bibinfo
  {pages} {144302} (\bibinfo {year} {2010})}\BibitemShut {NoStop}%
\bibitem [{\citenamefont {Iliev}\ \emph
  {et~al.}(1999{\natexlab{b}})\citenamefont {Iliev}, \citenamefont
  {Litvinchuk}, \citenamefont {Lee}, \citenamefont {Chen}, \citenamefont
  {Dezaneti}, \citenamefont {Chu}, \citenamefont {Ivanov}, \citenamefont
  {Abrashev},\ and\ \citenamefont {Popov}}]{Litvinchuk_PRB1999}%
  \BibitemOpen
  \bibfield  {author} {\bibinfo {author} {\bibfnamefont {M.~N.}\ \bibnamefont
  {Iliev}}, \bibinfo {author} {\bibfnamefont {A.~P.}\ \bibnamefont
  {Litvinchuk}}, \bibinfo {author} {\bibfnamefont {H.-G.}\ \bibnamefont {Lee}},
  \bibinfo {author} {\bibfnamefont {C.~L.}\ \bibnamefont {Chen}}, \bibinfo
  {author} {\bibfnamefont {M.~L.}\ \bibnamefont {Dezaneti}}, \bibinfo {author}
  {\bibfnamefont {C.~W.}\ \bibnamefont {Chu}}, \bibinfo {author} {\bibfnamefont
  {V.~G.}\ \bibnamefont {Ivanov}}, \bibinfo {author} {\bibfnamefont {M.~V.}\
  \bibnamefont {Abrashev}}, \ and\ \bibinfo {author} {\bibfnamefont {V.~N.}\
  \bibnamefont {Popov}},\ }\href {\doibase 10.1103/PhysRevB.59.364} {\bibfield
  {journal} {\bibinfo  {journal} {Phys. Rev. B}\ }\textbf {\bibinfo {volume}
  {59}},\ \bibinfo {pages} {364} (\bibinfo {year}
  {1999}{\natexlab{b}})}\BibitemShut {NoStop}%
\end{thebibliography}

%

\end{document}